\documentclass[12pt,preprint]{aastex}
\usepackage{rotating}
\usepackage{hyperref}
\usepackage[all]{hypcap}
\hypersetup{
    bookmarks=true,         % show bookmarks bar?
    unicode=false,          % non-Latin characters in Acrobat bookmarks
    pdftoolbar=true,        % show Acrobat toolbar?
    pdfmenubar=true,        % show Acrobat menu?
    pdffitwindow=false,     % window fit to page when opened
    pdfstartview={FitH},    % fits the width of the page to the window
    pdftitle={My title},    % title
    pdfauthor={Author},     % author
    pdfsubject={Subject},   % subject of the document
    pdfcreator={Creator},   % creator of the document
    pdfproducer={Producer}, % producer of the document
    pdfkeywords={keyword1} {key2} {key3}, % list of keywords
    pdfnewwindow=true,      % links in new PDF window
    colorlinks=true,        % false: boxed links; true: colored links
    linkcolor=blue,         % color of internal links (change box color with linkbordercolor)
    citecolor=blue,         % color of links to bibliography
    filecolor=magenta,      % color of file links
    urlcolor=cyan           % color of external links
}

\slugcomment{}

\shorttitle{Double Set Ribbon Solar Flare}
\shortauthors{Joshi et al.}

\begin{document}

%------------------------------------------------------------------------------------------

\title{Onset of a Large Ejective Solar Eruption from a Typical Coronal-Jet-Base Field Configuration}

%------------------------------------------------------------------------------------------

\author{Navin Chandra Joshi\altaffilmark{1}}
	\affil{School of Space Research, Kyung Hee University, Yongin, Gyeonggi-Do, 446-701, Korea; navin@khu.ac.kr, njoshi98@gmail.com}
\author{Alphonse C. Sterling\altaffilmark{2}}
	\affil{Heliophysics and Planetary Science Office, ZP13, Marshall Space Flight Center, Huntsville, AL 35812, USA}
\author{Ronald L. Moore\altaffilmark{2,3}}
	\affil{Heliophysics and Planetary Science Office, ZP13, Marshall Space Flight Center, Huntsville, AL 35812, USA}	
	\affil{Center for Space Plasma and Aeronomic Research (CSPAR), UAH, Huntsville, AL 35805, USA}
\author{Tetsuya Magara\altaffilmark{1,4}}
	\affil{School of Space Research, Kyung Hee University, Yongin, Gyeonggi-Do, 446-701, Korea}
	\affil{Department of Astronomy and Space Science, School of Space Research, Kyung Hee University, Yongin, Gyeonggi-Do, 446-701, Korea}
\author{Young-Jae Moon\altaffilmark{1,4}}
	\affil{School of Space Research, Kyung Hee University, Yongin, Gyeonggi-Do, 446-701, Korea}
	\affil{Department of Astronomy and Space Science, School of Space Research, Kyung Hee University, Yongin, Gyeonggi-Do, 446-701, Korea}

\altaffiltext{1}{School of Space Research, Kyung Hee University, Yongin, Gyeonggi-Do, 446-701, Korea}
\altaffiltext{2}{Heliophysics and Planetary Science Office, ZP13, Marshall Space Flight Center, Huntsville, AL 35812, USA}
\altaffiltext{3}{Center for Space Plasma and Aeronomic Research (CSPAR), UAH, Huntsville, AL 35805, USA}
\altaffiltext{4}{Department of Astronomy and Space Science, Kyung Hee University, Yongin, Gyeonggi-Do, 446-701, Korea}

%------------------------------------------------------------------------------------------

\begin{abstract}

Utilizing multiwavelength observations and magnetic field data from \textit{SDO}/AIA,
\textit{SDO}/HMI, \textit{GOES} and \textit{RHESSI}, we investigate a large-scale ejective solar
eruption of 2014 December 18 from active region NOAA 12241.  This event produced a distinctive
``three-ribbon'' flare, having  two parallel ribbons corresponding to the ribbons of a standard
two-ribbon flare, and a larger-scale third quasi-circular ribbon offset from the other two
ribbons. There are two components to this eruptive event. First, a flux rope forms above a
strong-field polarity-inversion line  and erupts and grows as the parallel ribbons turn on, grow,
and spread apart from that polarity-inversion line; this evolution is consistent with the
tether-cutting-reconnection mechanism for eruptions. Second, the eruption of the arcade that has
the erupting flux rope in its core undergoes magnetic reconnection  at the null point of a fan
dome that envelops the erupting arcade, resulting in formation of the quasi-circular ribbon; this
is consistent with the breakout-reconnection mechanism for eruptions.   We find  that the parallel
ribbons begin well before ($\sim$12~min) circular ribbon onset, indicating that tether-cutting
reconnection (or a non-ideal  MHD instability) initiated this event, rather than breakout
reconnection.  The overall setup for this large-scale  (circular-ribbon diameter $\sim$10$^5$~km)
eruption is  analogous to that of coronal jets (base size $\sim$10$^4$~km), many of which,
according to recent findings, result from eruptions of small-scale ``minifilaments.''  Thus these
findings confirm that eruptions of sheared-core magnetic arcades seated in fan-spine null-point
magnetic topology happen on a wide range of size scales on the Sun.

\end{abstract}

%------------------------------------------------------------------------------------------


%------------------------------------------------------------------------------------------

\section{Introduction}
\label{sec1}

Solar flares, the most violent explosions on the Sun, are vigorous magnetic energy releases
involving magnetic reconnection. The magnetic reconnection results in the formation of  flare
ribbons in the chromosphere and chromosphere-corona transition region, via the reconnection,  a
portion of the released  energy is converted into kinetic energy of accelerated charged particles
and another portion is converted directly into plasma thermal  energy \citep[see reviews by][and
references cited therein]{Benz08,Shibata11}. The reconnection starts in the corona, and
flare-ribbon formation occurs in the chromosphere immediately after  the onset of the
reconnection.  The ribbon emission results from  heating of the chromospheric plasma by the impact
of accelerated charged particles/electrons that stream downward from the reconnection region along
newly-reconnected field lines, and from heat conduction from the coronal plasma directly heated by
the reconnection  \citep[see review by][and references cited therein]{Fletcher01}. Solar flares can
be classified as ``confined" or ``eruptive/ejective," based respectively on whether they are or are
not produced in tandem with a coronal mass ejection (CME)\@. Several numerical simulation studies
investigating the magnetic configuration/topology and the initiation/triggering processes of solar
flares have been performed in two-dimensions (2D) and three-dimensions (3D)\citep[see reviews
by][and references cited therein]{Schmieder15,Janvier15}.

One of the well-accepted models for ejective flares is based on a 2D picture and is called
``CSHKP," because it is due to a combination of investigations by \cite{Carmichael64},
\cite{Sturrock66}, \cite{Hirayama74}, and \cite{Koop76}; this concept, sometimes augmented with
additional findings, is also sometimes referred to as ``the standard model for solar ejective
flares/eruptions.''   (In referring to this models for ejective flares/eruptions, we will use the
terms ``CSHKP'' and ``standard model'' interchangeably in the remainder of the text.) According to
this model, the flare reconnection occurs at a vertical current sheet underneath an erupting solar
filament/flux rope in the core of a magnetic  arcade. As the eruption and reconnection continue,
the erupting filament/flux rope becomes progressively more magnetically unleashed from the
photosphere and blows out the enveloping arcade  to make a CME, while newly-reconnected field below
the reconnection region forms strongly-emitting flare loops. Charged particles are accelerated at
the reconnection site, and move upward and downward from that site.  Those particles directed
upward are in reconnected field added to the  erupting flux rope. Downward-directed particles flow
along the reconnected field lines forming the flare loops, and the chromospheric material in the
flare-loop feet impacted by accelerated particles gets heated and appears as bright parallel flare
ribbons on either side of the magnetic polarity inversion line (PIL) of the pre-eruption arcade.
This kind of flare is commonly referred to as a  ``two-ribbon solar flare."  Flare
events have been observed showing various featurs of this model \citep[e.g.,][and references cited therein]{Tsuneta97,Shibata99,Benz08,Joshi13,Joshi16a}. This 2D CSHKP model alone
however does not explain several 3D characteristics of solar flares, including the sigmoid
structure of the erupting region, what initiates the eruption of the flux rope, the evolution of
the flare loops from being initially nearly aligned with the PIL to being nearly normal to it, and
other aspects. To explain these features, several 3D scenarios have been proposed
\citep{moore_et80,Shibata95,Moore01,Priest02,Shibata11,Janvier15}.

Apart from the parallel two-ribbon flares, flares with circular ribbons have also been observed and
studied \citep[e.g.,][]{Masson09,Su09,Reid12,Wang12,Sun13,Jiang13,wang_et14,Jiang14,Vemareddy14b,Joshi15}.
This type of ribbon is believed to result from reconnection at the magnetic null in a typical
fan-spine magnetic configuration.  This configuration consists of inner and outer fan shells, with
the inner shell being a closed  dome of loops having one foot in a minority-magnetic-polarity 
region, and the outer shell being a dominant-magnetic-polarity canopy of far-reaching field
sheathing the dome and  forming a magnetic spine field that is open or connects to the solar
surface at a remote location. These inner and outer fan regions are separated by a separatrix
layer, resulting in a magnetic null between the top of the closed inner  shell and the open outer
shell (see Figure~1 of \cite{Pariat09} and \cite{Sun13}). As we will discuss further below,
reconnection at the null plausibly results in the circular ribbon.  Possibilities for driving
magnetic reconnection at this null include eruption of an underlying filament/flux rope from low
along the PIL of the closed-field dome \citep[]{Sun13,Jiang13,Jiang14,Joshi15,Joshi16b}; and shearing of the
dome field lines, making the dome field press on the null  \citep[e.g.,][]{Vemareddy14b}. Some
studies also suggest that quasi-separatrix-layer reconnection in the top of the separatrix layer
also can make circular-ribbon flares \citep[e.g.,][]{Reid12}. There are only a few papers that deal
with  simulations of these kinds of eruptions and null-point reconnections
\citep{Masson09,Torok09,Pariat09,Jiang13}.

From observations, we know that the eruption of a filament, sigmoid, and/or flux-rope  structure
plays a crucial role in driving solar flares.  Several models have been suggested for this aspect
of  solar eruptions and flares. In the ``tether-cutting" scenario
\citep{moore_et80,Sturrock84,Moore01}, it was suggested that, initially, shearing and converging 
photospheric flow brings opposite-polarity legs of a sheared magnetic arcade into contact at a PIL
to initiate runaway tether-cutting reconnection. This reconnection forms and/or grows a flux rope
above the reconnection location, and this flux rope immediately starts erupting outward.  As it
does so, the region of the magnetic reconnection above the PIL and below the outward-moving flux
rope continues to grow larger and  move higher; this is the flare reconnection in the above
description of the CSHKP model. Observational events have been observed and interpreted using this model \citep[e.g.][and references cited therein]{Liu12,Liu13,Chen14,Joshi14,Xue17}. 

Another eruption model, called ``magnetic breakout" \citep{Antiochos99}, assumes an initial
quadrupolar magnetic configuration, where a flux rope erupts from the PIL of a ``core'' sheared
magnetic  arcade embedded in an opposite-direction larger arcade.  The model argues that
reconnection at the magnetic null above the sheared arcade initiates the eruption of the sheared
arcade; the strongest flare reconnection then occurs at a current sheet at, again, the same
location as in the CSHKP model: above the PIL of the sheared core field and below the  erupting
flux rope. Various observational features of this model have been observed in observational events \citep[e.g.][and references cited therein]{Aulanier00,Sterling01,Sterling04,Aurass13,Chen16}.

Other ideas have also been put forth for the onset of the magnetic eruptions that make flares and
CMEs, including the ideal magnetohydrodynamic instabilities known as the ``kink
instability"\citep{Torok04,Torok05} and the ``torus instability"\citep{Kliem06}. Also, recently
there have been other proposed 3D models based on both observations and simulations, such  as  the
``standard model in three dimension" \citep{Aulanier12,Janvier14,Li15,Gou16,Dudik16}. This MHD model includes the eruption of a torus-unstable magnetic
flux rope via the torus instability, and reconnection at a current sheet that forms underneath the erupting flux rope.  The current layers are regions of 
high magnetic field gradient, known as ``quasi-separatrix layers” \citep[QSLs;][]{Priest95,Demoulin96}. The
slipping-reconnection concept has also been proposed as a mechanism to initiate solar 
flares/eruptions \citep{Aulanier06,Dudik14}. Moreover, the slipping reconnection itself has been proposed to be the tether-cutting mechanism \citep{Dudik16}. Models such as the ``slipping reconnection" model \citep[e.g.][]{Dudik16} or the ``Standard solar flare model
in 3D," \citep[e.g.][]{Aulanier12} incorporate the 2D CSHKP model in a 3D view.

The field configurations and initiation of three-ribbon flares, with two parallel ribbons and  a
third circular ribbon occurring in the same flux-rope-eruption episode, have not been investigated 
extensively. In one recent such study, \cite{Joshi15} discussed the magnetic topology and
triggering of such a three-ribbon flare; they suggested a possible scenario for the event they
observed, and recently \cite{Carley16} presented  support for that scenario based on solar radio
data.  In the present work, we present observations of a different example of a similar
three-ribbon-flare eruption.   It is an ejective eruption of a sigmoid flux rope, and occurred on
2014 December 18 in a  large-scale fan-spine type magnetic configuration. Our main aim in this
study is to determine how the  eruption was initiated and how it produced the three-ribbon flare. 
We will also discuss similar  eruptions that occur on a much smaller size  scale in the form of
coronal jets; recently it has been recognized that many such jets result from small-scale
``minifilament'' eruptions, which appear to be scaled-down versions of larger  sheared-core arcade
eruptions that make flares and CMEs \citep{Sterling15}.  The magnetic  geometry we describe for our
large-scale three-ribbon flare is also present in the  jets produced by minifilament eruptions, and
those jets also have a similar pattern of three flare ribbons.  

We construct this paper in the following sections. Section~\ref{sec2} explains about the observational data sets used. Source region location information and overlying magnetic topology are discussed in Section~\ref{sec3}. Multiwavelength analysis are presented in Section~\ref{sec4}, with GOES and RHESSI flux temporal evolution (Section~\ref{sec4.1}), parallel ribbon formation dynamics (Section~\ref{sec4.2}), circular ribbon formation dynamics (Section~\ref{sec4.3}), discussion about ribbons formation timings (Section~\ref{sec4.4}), and associated CME (Section~\ref{sec4.5}). A comparision with the coronal jet topology is presented in Section~\ref{sec5}. The main results and their discussions are presented in Section~\ref{sec6}.

%------------------------------------------------------------------------------------------

\section{Instrumentation} 
\label{sec2} 

We use data from several  space-based instruments:
\textit{SDO}/AIA and HMI, \textit{GOES} and \textit{RHESSI}\@. The Atmospheric Imaging Assembly
\citep[AIA;][]{Lem12} is an instrument on board the {\it Solar Dynamics Observatory}
(\textit{SDO}) that observes the Sun in visible, ultraviolet (UV), and extreme ultraviolet (EUV)
wavelengths. For the present work we have examined AIA observations at all available EUV
channels:  304, 131, 171, 193, 211, 335, and 94~\AA, and also in UV at 1600 and 1700~\AA, and in visible
light at 4500~\AA\@. AIA has temporal cadences of 12~s for EUV and 24~s for UV, and
pixel size of $\rm 0.6\arcsec$. The Helioseismic and Magnetic Imager \citep[HMI;][]{Schou12} is
another instrument on board \textit{SDO}. It provides line-of-sight magnetic field 
maps of the whole face of the Sun with a temporal cadence of 45~s and with pixel size of $\rm 0.5\arcsec$.
For observing the X-ray-coronal and footpoint sources during the flare we use data from the {\it
Reuven Ramaty High-Energy Solar Spectroscopic Imager} \citep[\textit{RHESSI};][]{Lin02}, where we
used the PIXON algorithm for image reconstruction with an integration time of around 20~s.

%------------------------------------------------------------------------------------------

\section{Source Region Location, Morphology and Magnetic Field Structure} 
\label{sec3}

\begin{figure}
\vspace*{-8cm}
\centerline{
	\hspace*{0.\textwidth}
	\includegraphics[width=1.3\textwidth,clip=]{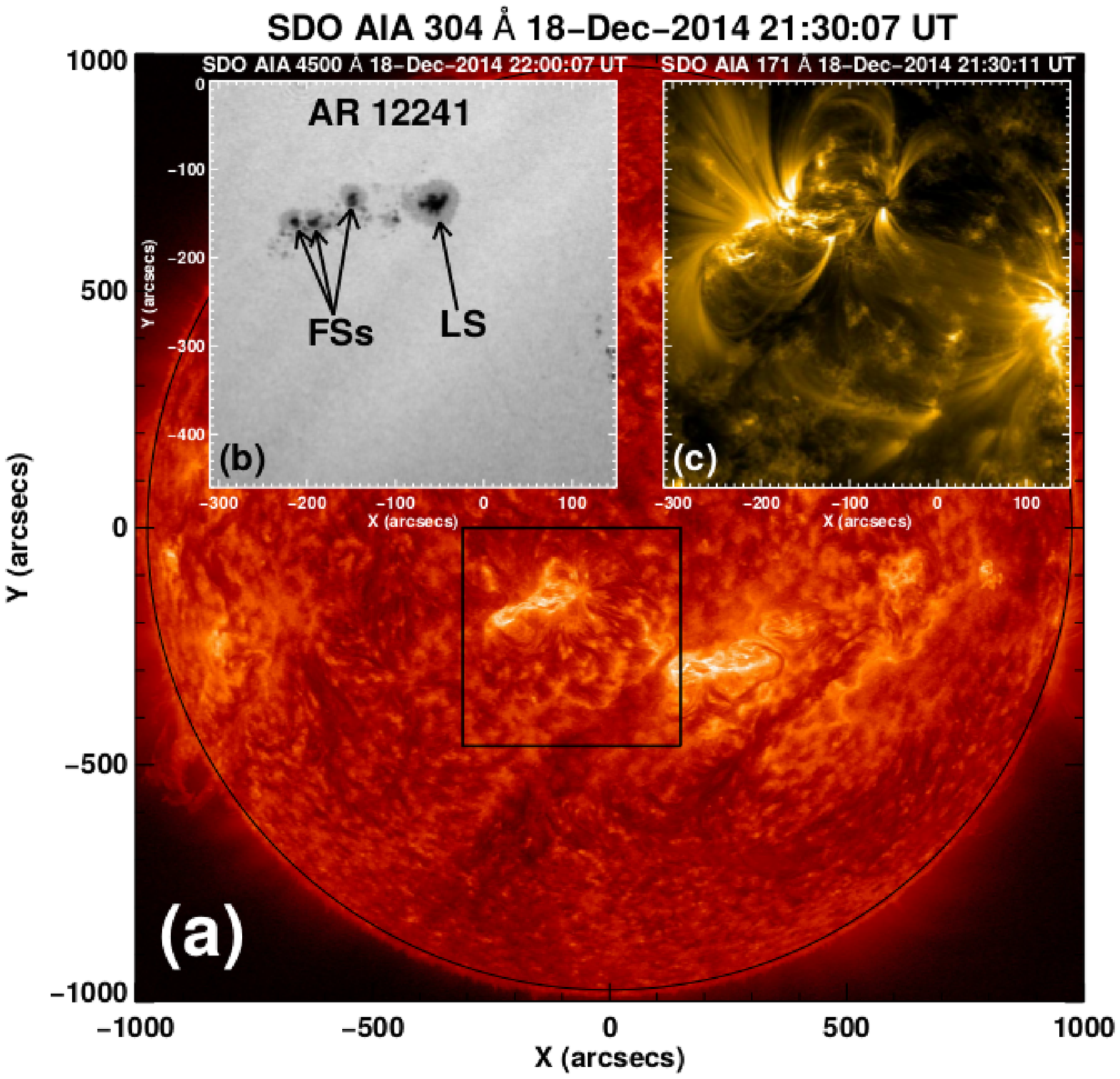}
	}
\vspace*{-6cm}
\caption{(a)(Background image) \textit{SDO}/AIA 304 \AA\ full disk image at 21:30:07 UT 
on 2014 December 18. The black box shows the active region location and the area of flare 
activity.  Inserts are \textit{SDO}/AIA 4500 (b) and 171 \AA\ (c) zoomed images, corresponding 
to the black box shown in (a). The leading sunspot (LS) and following 
sunspots (FSs) are indicated in panel (b).}
\label{fig1}
\end{figure}

Figure~\ref{fig1} is a montage of \textit{SDO}/AIA 304, 171 and 4500 \AA\ images showing an
overview of the active region (AR) that was the source of the event, NOAA AR 12241. The background
image is a \textit{SDO}/AIA 304 \AA\ full disk image from 21:30:07 UT on 2014 December 18
(Figure~\ref{fig1}(a)). The black box in Figure~\ref{fig1}(a) shows the location of AR 12241, with
closeup images in \textit{SDO}/AIA 4500 \AA\ and 171 \AA\ respectively in Figures~\ref{fig1}(b)
and \ref{fig1}(c).  These images show that AR 12241 is located near disk center, and the 
\textit{SDO}/AIA 4500 \AA\ image shows that the AR consists a group of several sunspots,
with one leading sunspot and  three relatively small following sunspots (arrows in
Figure~\ref{fig1}(b)). The magnetic class of the active region was $\alpha$$\beta$$\gamma$ on 2014
December 18.

\begin{figure}
\vspace*{-8cm}
\centerline{
	\hspace*{0.0\textwidth}
	\includegraphics[width=1.3\textwidth,clip=]{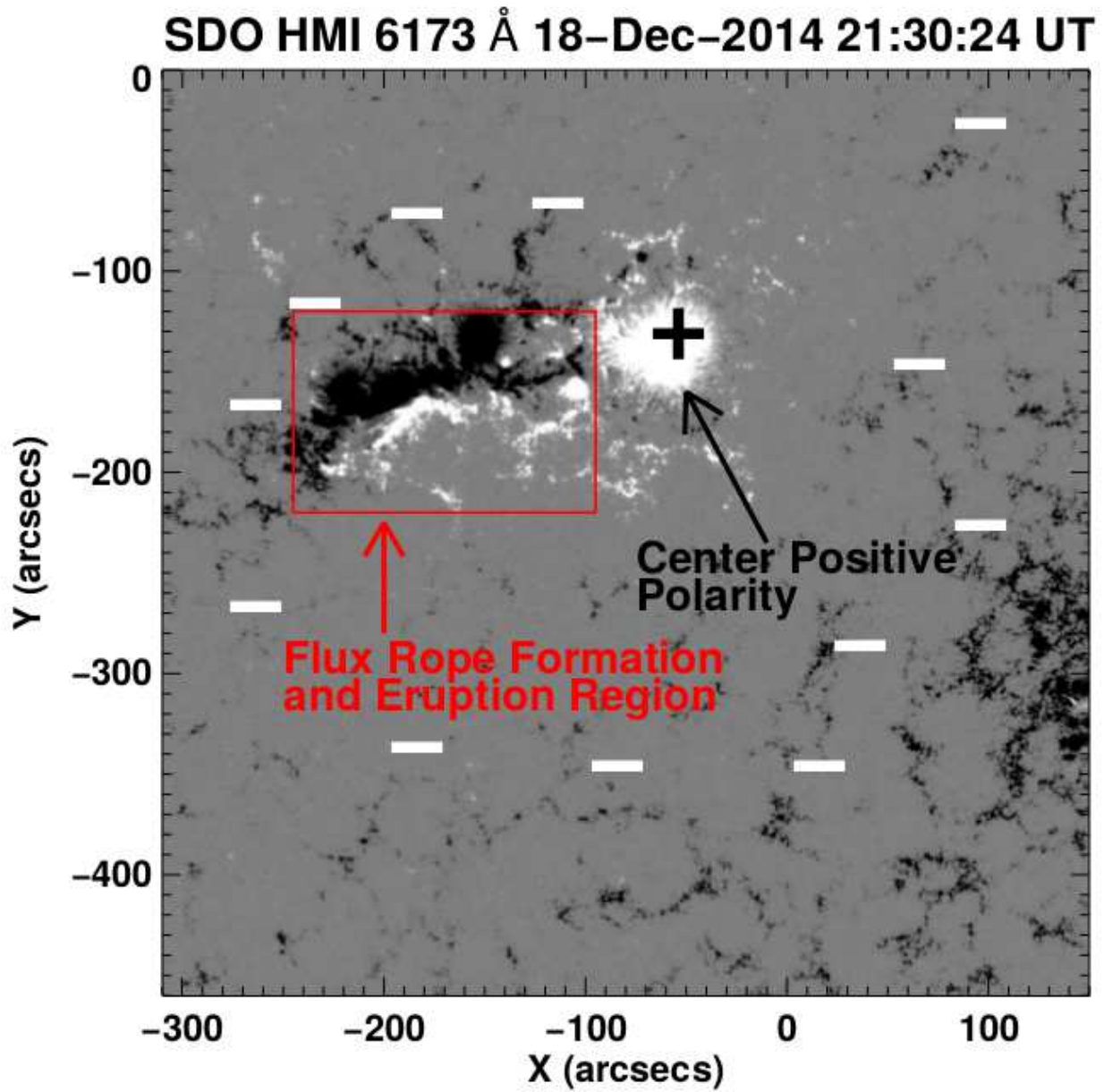}
	}
\vspace*{-6cm}
\caption{\textit{SDO}/HMI line-of-sight photospheric magnetogram at 21:30:24 UT on 2014 December 18. The area of this image corresponds to the black box in Figure~\ref{fig1}(a). The red box represents the site of flux-rope formation and eruption.}
\label{fig2}
\end{figure}

Figure~\ref{fig2} shows the line-of-sight photospheric magnetogram corresponding to the area shown
in the black box of Figure~\ref{fig1}(a). This magnetogram is from 21:30:24 UT on 2014 December 18,
and shows that the active region is bipolar with much of its positive-polarity magnetic flux in the
large leading sunspot and much of its negative flux in the three following sunspots, and has a long
PIL between its positive and negative flux domains.  Also marked in the figure is an extensive
region of negative polarity that surrounds the positive flux domain region. This photospheric
magnetic distribution suggests that there should be a large-scale fan-spine type configuration with
a shell of open or far-reaching field  outside of the AR that is rooted in the surrounding
negative-polarity region and extending far above and away from the AR; thus the AR would be inside
of this shell, and so will call this shell the ``outer shell.''  There is also a portion of the fan
field inside of that outer shell, i.e.\ an ``inner shell,'' which is an  anemone-shaped ring of
field loops having one foot in the positive flux domain and the other foot in the surrounding
negative domain.  As we will see shortly, the strip along the PIL in the red box is the site of
formation of a flux rope that erupts; it is that eruption and its interaction with the shells of
the fan that produce the three flare ribbons. 

\begin{figure}
\vspace*{-8cm}
\centerline{
	\hspace*{0.0\textwidth}
	\includegraphics[width=1.2\textwidth,clip=]{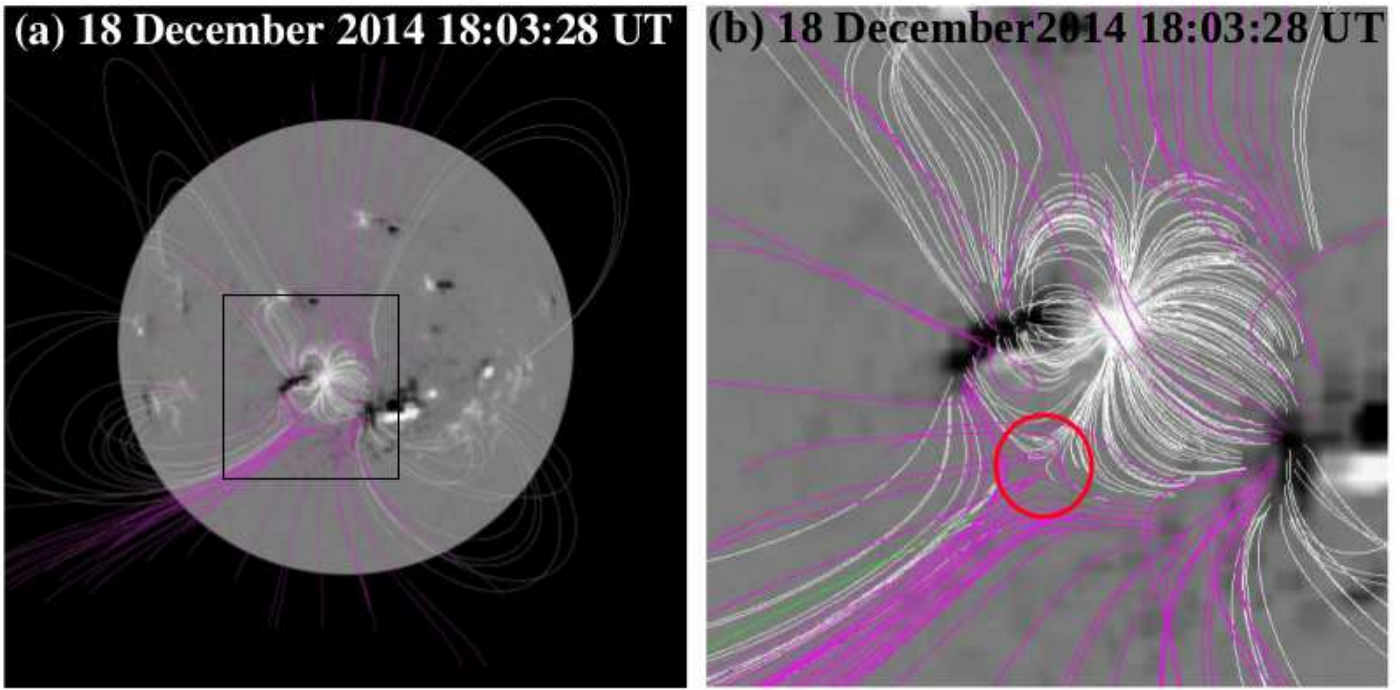}
	}
\vspace*{-8cm} 
\caption{(a) PFSS magnetic field extrapolation showing a full disk view of the coronal magnetic field structure over the active region. (b) A zoomed view corresponding to the black box in (a).  The 
white and pink lines represent closed and open field lines rooted in and around the active 
region, respectively. The magnetic null is inside the red circle in (b).} 
\label{fig3}
\end{figure}

In Figure~\ref{fig3} we confirm the existence of a huge fan-spine structure using a potential-field
source-surface \citep[PFSS,][]{Schrijver03} extrapolation in the neighborhood of the AR, where we
accessed the PFSS package with the Interactive Data Language (IDL) using the Solar Software (SSW)
package \citep{freeland_et98}.  Figure~\ref{fig3}(a) shows a full disk magnetogram from 18:03:28 UT on
2014 December 18 with the extrapolated potential field lines overplotted, and Figure~\ref{fig3}(b)
shows a closeup of the active region corresponding to the region of the black box in
Figure~\ref{fig3}(a).  The closed field lines (shown in white) are the field lines of the inner fan
dome, while the open field lines (pink) represent the field lines of the outer shell of the fan dome.
The inner (closed) and outer (open) fan-dome fields are expected to be separated by a separatrix layer
in 3D\@. As we will show below, this configuration is well-matched with the \textit{SDO}/AIA
observations (see Sections~\ref{sec3}, and~\ref{sec5}). Later (Section~\ref{sec5}), we will show that this 
configuration is similar to a typical coronal-jet-base field configuration with an inferred fan-dome
field topology in the corona \citep{Sterling15}. 

%------------------------------------------------------------------------------------------

\section{Multiwavelength Analysis of the Ejective Eruption and Flare Dynamics}
\label{sec4}

%------------------------------------------------------------------------------------------

\subsection{\textit{GOES} and \textit{RHESSI} X-ray Intensity Profiles}
\label{sec4.1}

\begin{figure}
\vspace*{-7cm}
\centerline{
	\hspace*{-0.0\textwidth}
	\includegraphics[width=1.4\textwidth,clip=]{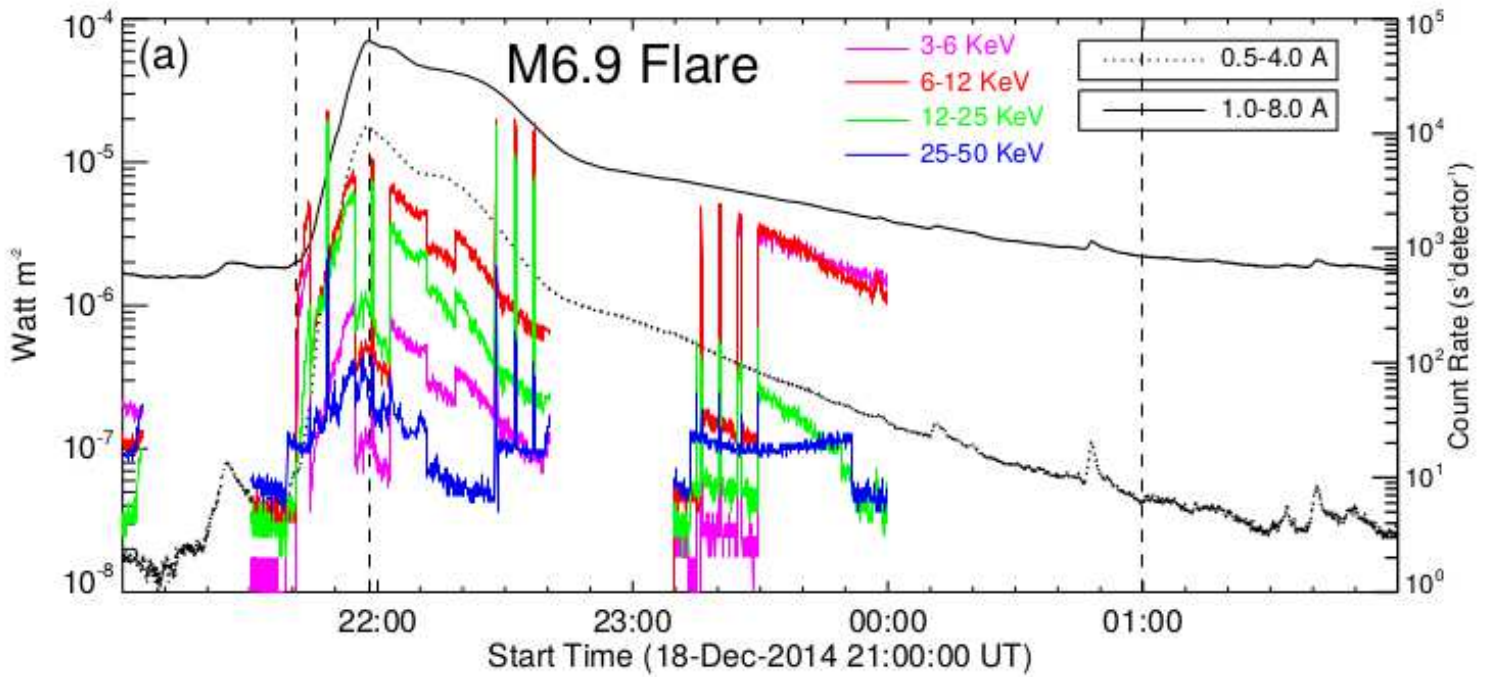}
	}
\vspace*{-10.5cm}
\caption{\textit{GOES} and \textit{RHESSI} X-ray flux time profiles from 21:00 UT on 2014 December 18 to 03:00 UT on 2014 December 19. The flare's start ($\sim$21:41 UT), peak ($\sim$21:58 UT) and very late ($\sim$1:00 UT) times are indicated 
by the three dashed lines from left to right, respectively.}
\label{fig4}
\end{figure}

Figure~\ref{fig4} shows the {\it Geostationary Operational Environmental Satellite} (\textit{GOES})
and \textit{RHESSI} X-ray time profiles of the flare. The black solid and dashed curves are the
\textit{GOES} X-ray profiles for 1--8 \AA\ and 0.5--4 \AA\, respectively,  and the \textit{RHESSI}
X-ray profiles at 3-6 keV (pink), 6-12 keV (red), 12-25 keV (green), and 25-50 (blue) keV are
overplotted. This flare was class M6.9 on the \textit{GOES} X-ray scale. From  the \textit{GOES}
profiles we see that the flare started at $\sim$21:41 UT and peaked at $\sim$21:58 UT,
and had an extended decay phase that continues until $\sim$02:00 UT on 2014 December~19. The overall
variation of the \textit{RHESSI} X-ray profiles matches with that of the \textit{GOES} profiles.

%------------------------------------------------------------------------------------------

\subsection{Two-Ribbon Component: Standard Reconnection and Parallel Ribbons Formation}
\label{sec4.2}

\begin{figure}
\vspace*{-8cm}
\centerline{
	\hspace*{0.0\textwidth}
	\includegraphics[width=1.3\textwidth,clip=]{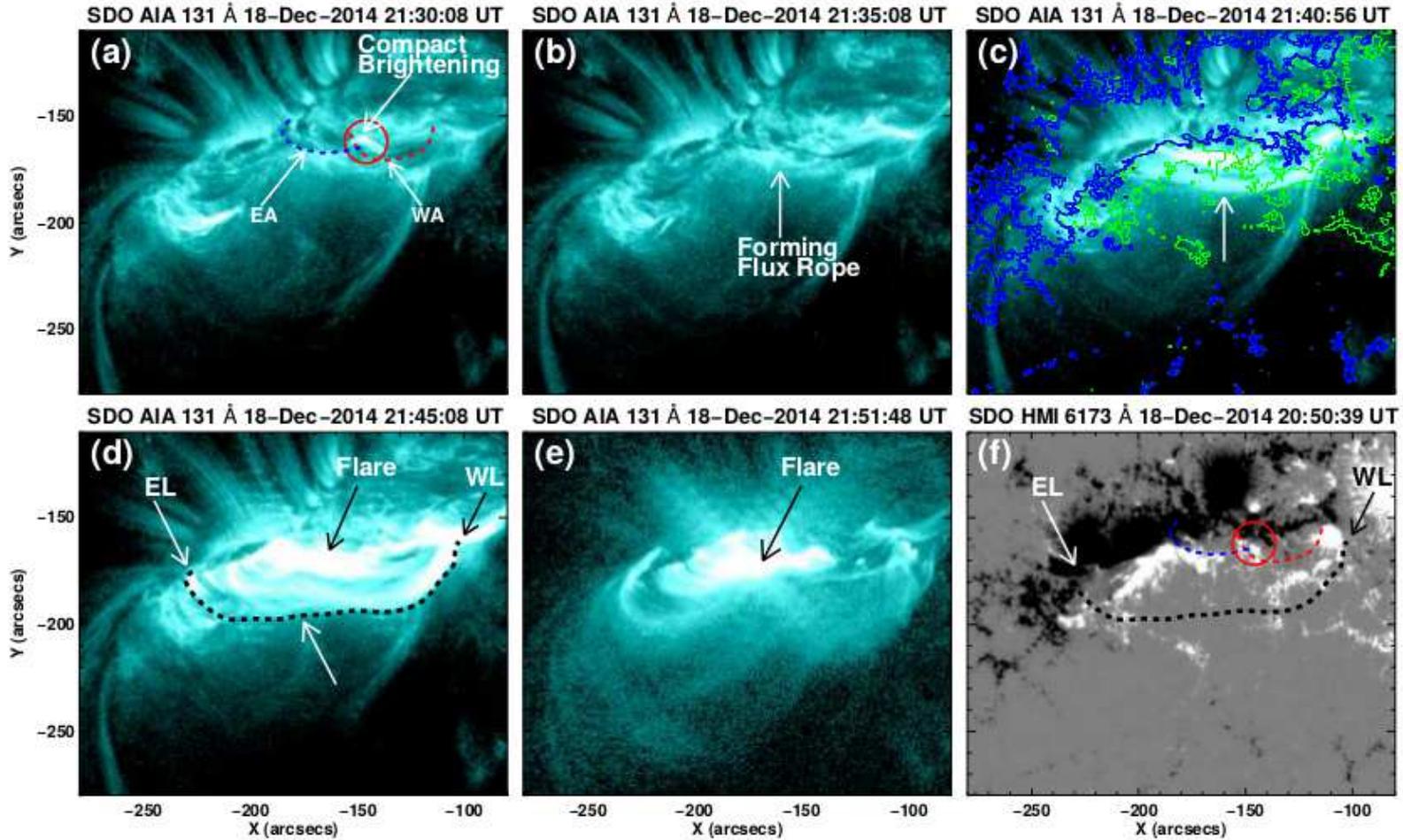}
	}
\vspace*{-7.5cm}
\caption{((a)--(e)) \textit{SDO}/AIA 131 \AA\ images showing the evolution of the two-ribbon component of the eruptive event, with formation and eruption of the flux rope and standard flare reconnection over $\sim$21:30 UT to $\sim$21:51 UT on 2014 December 18. (f) \textit{SDO}/HMI line-of-sight photospheric magnetogram at 20:50:39 UT on 2014 December 18. The dashed red and blue curved lines in panels (a) and (f) represent the eastern (EA) and western (WA) loops of the pre-eruption sheared arcade, respectively. The red circle in panels (a) and (f) represents 
the location of an initial compact brightening region in (a). The curved dotted black line in panels (d) and (f) show the front part of the moving flux rope at 21:45:08 UT\@. In panel (c) the blue and green contours show the photospheric magnetic field contours for negative- and positive-polarity field, respectively, with
contour levels of $\pm40,\pm50, \rm{and } \pm100$ Gauss.} 
\label{fig5}
\end{figure}

The event consists of two different aspects, or ``components.'' One component is a standard
sheared-core-arcade flux-rope eruption, including the standard CSHKP-type flare reconnection and
parallel ribbon formation discussed above, all occurring inside of (and perhaps initiated independently
of) the large-scale fan-spine structure. Figures~\ref{fig5}((a)--(e)) present a sequence of
\textit{SDO}/AIA 131~\AA\ images during the time of dynamic evolution of the flux rope, including its
formation, eruption, and triggering of CSHKP-type flare reconnection. The first indication of the flare
onset in these images seems to be at $\sim$21:30 UT, with a compact brightening at the location of the 
red circle in Figure~\ref{fig5}(a). This brightening may be due to tether-cutting reconnection
occurring between low-lying sheared arcade loops with the  footpoints of their legs anchored in 
opposite-polarity footpoint regions.  We can identify candidates for these loops as dark arch-like
structures on the eastern and western sides of the red circle; they  are shown as red- and blue-dashed
curves in Figure~\ref{fig5}(a), and respectively  labeled ``WA'' and ``EA'' for western and eastern
loops of a highly-sheared arcade.   We further examine the anchoring of the arcade loops as well as the
location of the compact brightening by comparing the AIA 131 \AA\ images with \textit{SDO}/HMI
photospheric magnetogram, by overplotting WA and EA of Figure~\ref{fig5}(a) onto the magnetogram of
Figure~\ref{fig5}(f).  This shows that the far ends of the WA and EA arcade loops are anchored in
negative and positive polarity regions, respectively. The location of the initial compact brightening
region (red circles in Figs.~\ref{fig5}(a) and~\ref{fig5}(f)) is near the junction of the two arcades. 

Inspection of the \textit{SDO}/HMI photospheric magnetic field movie (see animation accompanying
Figure~\ref{fig5}) of the area shows rapid evolution, with continuous flux
cancellation and shearing motions; in particular, the red-circled area in Figure~\ref{fig5}(f)
shows cancelation occurring in the hours prior to the onset of flaring, with an episode of
cancelation between the positive-polarity patch with adjacent negative-polarity flux
starting at about 20:15~UT\@. This rapid flux cancelation
is a strong candidate for driving pre-eruption tether-cutting reconnection that could build
a pre-eruption core flux rope.  That is, this pre-eruption tether-cutting reconnection is apparently 
driven by flux cancelation resulting from evolutionary photospheric motions.

Soon after the initial brightening, we observe the formation and appearance of long thread-like
structures over the flux-cancellation region (Figures~\ref{fig5}(b)--(e)). These are perhaps best
visible in 131~\AA\ and 94~\AA\ wavelengths, from about 21:35~UT (or even earlier in 94~\AA),
where they appear in emission; they are faintly visible in  193~\AA\ images, but more  difficult
to see or invisible at this early stage (when they are low down in the core region) in
other wavelengths.  These newly-formed long structures are consistent with them being a component
of the flux rope expected to result from tether-cutting reconnection below.  Thus, similar to some
other cases \citep[e.g.,][]{cheng_et13,sterling_et14}, the erupting flux rope tends to be visible
in (some) hotter AIA EUV channels, but faint or invisible in the cooler AIA  channels.  

In this case, although a cool-material filament is clearly visible along  the PIL, there is no
clear evidence that any of this filament material is ejected along with the hotter flux rope. (A
strand of this filament is ejected from the west side of the filament at $\sim$21:51~UT, but this
is a secondary aspect  of the flux-rope eruption, in the sense that it occurs well after the two
parallel ribbons start to  separate; we do not consider here whether this strand ejection is a
connected to the null-point  reconnection (Section~\ref{sec4.3}) that begins at about the same
time.) Ribbon-like brightenings from the eruption start parallel to the filament along either side
of the filament; these features are perhaps most apparent in the 304~\AA\ images (not included
here), but they can also be seen in the movie accompanying Figure~\ref{fig5}(a).  This suggests
that the suspected tether-cutting reconnection occurred among fields overlying the filament
\citep[e.g.,][]{moore_et07}, and therefore the  flux-rope formed and was expelled from a location
above the filament, leaving the filament inside the flare arcade.

In Figure~\ref{fig5}(d) we trace the observed front of the newly-formed flux rope with a  dotted
black line, and in Figure~\ref{fig5}(f)) we overplot that trace onto an \textit{SDO}/HMI
photospheric magnetogram. This shows that the eastward leg of the observed  flux rope is anchored
in a negative-polarity region, and the westward leg is anchored in a  positive-polarity region. 
This configuration supports that this observed outward-moving  bright feature is indeed a flux
rope that forms from runaway tether-cutting reconnection early in the eruption.  (By ``runaway
tether cutting,'' we mean a situation whereby the tether-cutting reconnection promotes further
tether-cutting reconnection through a positive-feedback  process for an extended period of time.)
We also observe flare brightening along the PIL location below the erupting flux rope in all AIA
EUV wavelengths as the flux rope moves outward (see Figures~\ref{fig5}(d) and (e), and
accompanying video). This flaring is expected as the legs of the arcade enveloping the erupting
flux rope partake in the CSHKP-type reconnection.

\begin{figure}
\vspace*{-8cm}
\centerline{
	\hspace*{0.0\textwidth}
	\includegraphics[width=1.3\textwidth,clip=]{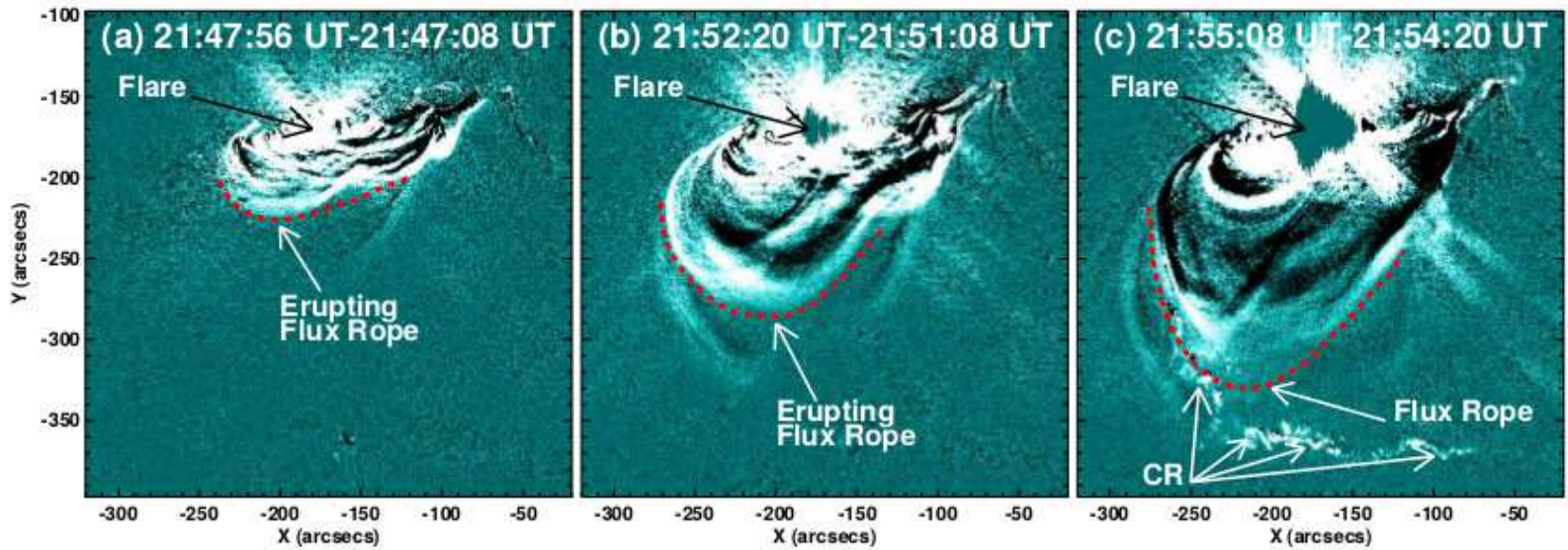}
	}
\vspace*{-10cm}
\caption{\textit{SDO}/AIA 131 \AA\ running-difference images showing the erupting flux rope, the 
two-ribbon-flare ribbons, and the circular ribbon (CR)\@. The red dashed line shows the 
leading edge of the erupting flux rope.} 
\label{fig6}
\end{figure}

\begin{figure}
\vspace*{-8cm}
\centerline{
	\hspace*{0.0\textwidth}
	\includegraphics[width=1.3\textwidth,clip=]{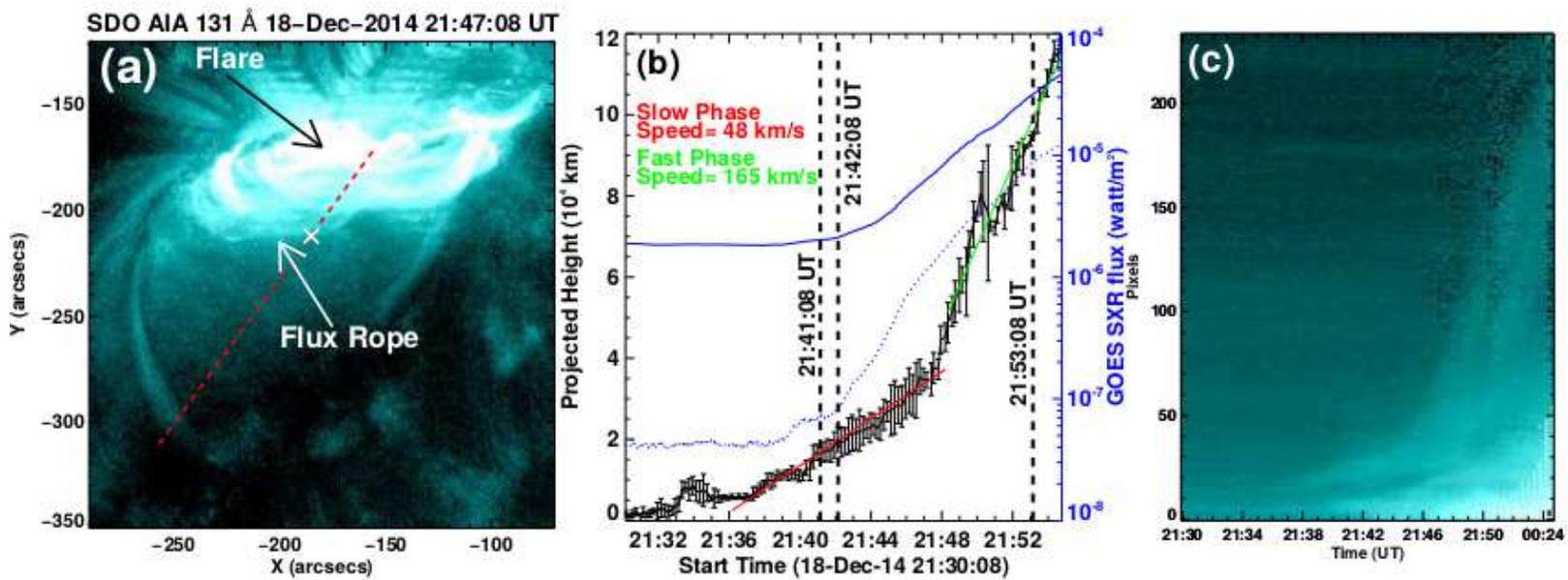}
	}
\vspace*{-8.5cm}
\caption{Left panel: \textit{SDO}/AIA 131 \AA\ image at 21:47:08 UT on 2014 December 18. The red 
dashed line is the path along which the height measurements and distance-time map are made. Middle panel: 
height-time profile of the leading edge of the erupting flux rope. The \textit{GOES} X-ray profiles at 1--8 (blue 
solid curve) and 0.5-4 (blue dotted curve) \AA\ channels are also overplotted. The speeds are 
from the red and green linear fits to the height-time data points. The "X" in panel (a) represents the location where the flux rope is being measured at the time of this image. The vertical dashed lines from left to right represent the start time of flare, first appearence of parelle ribbons and first appearence of circular ribbon. Right panel: Time-distance map of the eruption long the red dashed line of (a).} 
\label{fig7}
\end{figure}

Figure~\ref{fig6} shows the outward propagation of the flux rope in \textit{SDO}/AIA 131~\AA\ 
running-difference images, where we highlight the approximate front of the flux rope by the  dashed
red lines. We can gain insight into the temporal relation between the erupting flux rope, the flare
brightening, and the formation of the ribbons by comparing the height-time profile of the erupting
flux rope's leading edge  with the \textit{GOES} X-ray profiles (Figure~\ref{fig7}). 

The left panel of Figure~\ref{fig7} shows an \textit{SDO}/AIA 131 \AA\ image at 21:47:08 UT and
the  red dashed line along which the projected height of the erupting flux rope has been measured.
The height-time profile is shown in the right panel of Figure~\ref{fig7}. To determine the
height-time data points, we tracked the leading edge of the erupting  flux rope along the red
dashed line in Figure~\ref{fig7}(a), where the bottom-most point of the line is used as a reference
point for the height measurements. To estimate the uncertainty in the  height-time locations we
repeated the same analysis three times, and the average of three  values has been used as a final
value of height-time measurements and the error bars are the 1$\sigma$ standard deviations
resulting from those three repeated measurements. The speeds are  calculated using linear fits to
the different time intervals. The height-time profile reveals a two-phase evolution of the
eruption: The first phase is during $\sim$21:36 UT to  $\sim$21:48 UT, with an average speed of
around $\rm 48~km~s~^{-1}$, while the
second phase is from $\sim$21:48 UT to $\sim$21:55 UT with a relatively higher average speed of
around $\rm 165~km~s^{-1}$. The first
and second phases are shown by the red and green fitted lines in Figure~\ref{fig7}(b),
respectively. We also  overplotted the \textit{GOES} 1-8 and 0.5-4 \AA\ flux profiles with the
height-time profile (shown by blue thick and dotted lines, respectively, in Figure~\ref{fig7}(b)).
It can be seen that the \textit{GOES} X-ray flux enhancement is well matched with the height-time
profile, suggesting that the flux rope's upward motion is correlated with runaway tether-cutting
reconnection underneath it.  Two-step rise profiles (slow-rise and fast-rise phases) have commonly
been seen for filament eruptions 
\citep[e.g.,][]{tandberg-hanssen_et80,sterling_et05,mccauley_et15}, and in some cases accelerations
in the filament trajectory have been also found to correspond to peaks in brightenings in
\textit{GOES} fluxes  \citep[e.g.][]{sterling_et07,sterling_et11}.  Thus, the flux-rope eruption
behavior here is in agreement with those observations of erupting filaments (which are believed to
be cool material riding on erupting flux ropes). The time-distance map of the eruption have also been shown in Figure~\ref{fig7}(c).

\begin{figure}
\vspace*{-8cm}
\centerline{
	\hspace*{0.0\textwidth}
	\includegraphics[width=1.4\textwidth,clip=]{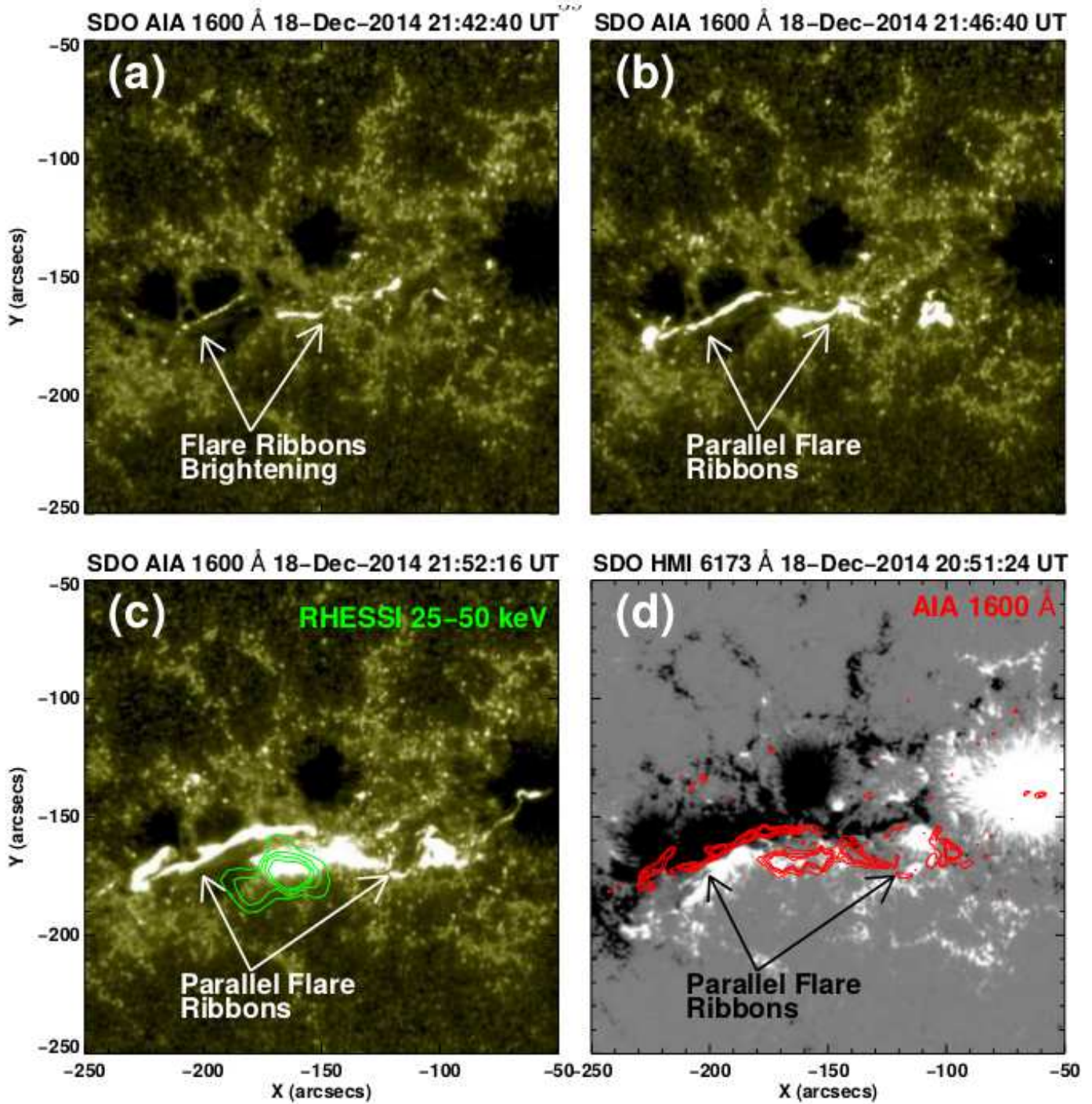}
	}
\vspace*{-5.5cm}
\caption{((a)--(c)) \textit{SDO}/AIA 1600 \AA\ image sequence, showing the formation of standard parallel ribbons during the first stage of the eruption event. (d) \textit{SDO}/HMI line-of-sight photospheric magnetogram at 20:51:24 UT. The overplotted red contours are of \textit{SDO}/AIA 1600 \AA\ intensity, showing
the parallel ribbons at 21:52:16 UT (the time of panel (c)). The contour levels are 5\%, 10\%, 20\%, and 30\% of the peak intensity. The green contours in panel (c) image the \textit{RHESSI} X-ray intensity in the energy range 25-50 keV, generated with the PIXON algorithm. The contours 
levels are 20\%, 30\%, 40\%, and 50\% of the peak intensity, and the 
integration time is 20~s.}
\label{fig8}
\end{figure}

The morphology of the parallel ribbons formation can be seen in the \textit{SDO}/AIA 1600 \AA\
observations (Figure~\ref{fig8}). The initial ribbon  brightening is observed over
$\sim$21:41---$\sim$21:42 UT (Figure~\ref{fig8}(a)).  Early in  the evolution of the flux rope,
e.g.\ at about 21:47~UT (Figure~\ref{fig8}(b)), we see multiple ribbons, broadly  consistent with
the four-ribbon configuration expected during the earliest stages of the  tether-cutting scenario
\citep[see top-right panel of Fig.~1 in][]{Moore01}.  Later (e.g.,  22:10~UT), these ribbons evolve
into the standard parallel ribbons (see Figure~\ref{fig10}(d)) along both sides of the PIL; the
overall dynamics can be seen in the \textit{SDO}/AIA 1600 \AA\ movie (see animation  accompanying
Figure~\ref{fig10}). Contours of the standard parallel ribbons observed in \textit{SDO}/AIA 1600
\AA\ are overplotted on line-of-sight HMI magnetogram in Figure~\ref{fig8}(d), showing that the
northern and southern parallel ribbons are located in negative and positive polarity regions,
respectively. The ribbons extend in the east-west direction and move apart in the north-south
direction as the flux rope moves outward. Moreover, we also observed RHESSI 25--50 keV X-ray source on one of the two ribbons (see green contours in Figure~\ref{fig8}(c)). These observations are consistent with the  standard
flare model, with runaway tether-cutting reconnection occurring below the erupting flux rope and
with charged particles  accelerated at the reconnection site and flowing downward along the
magnetic field  lines and striking the lower chromosphere to emit hard X-rays and heat the flare
ribbons.  

Figure~\ref{fig9} presents a  schematic based on the observations in the various AIA EUV and
1600~\AA\ channels and these  discussions. We emphasize that all of these early-stage dynamics
(erupting flux rope and two-ribbon flare)  are occurring inside of the northeastern lobe of the
huge fan-dome magnetic-field  configuration (see red box in Figure~\ref{fig2}).

%------------------------------------------------------------------------------------------

\subsection{Circular-Ribbon Component: Null point Reconnection and Circular Ribbon Formation}
\label{sec4.3}

This component of the eruption is a consequence of the null-point reconnection, which results in 
the circular ribbon. It occurs because of the interaction of the erupting flux rope and the arcade
containing it (where that arcade is  the northeast side of the inner fan field prior to its 
eruption) described in
Section~\ref{sec4.2} with the null of the fan-spine magnetic structure. The formation and dynamics 
of the circular ribbon can be understood by
analyzing the \textit{SDO}/AIA observations in different wavelengths (Figures~\ref{fig10}
and~\ref{fig11}). 

Figure~\ref{fig10} shows a sequence of selected \textit{SDO}/AIA 1600 \AA\ images over
$\sim$21:53 UT to $\sim$22:10 UT\@. The first appearance of the circular ribbon is between
21:52~UT and 21:53~UT, starting near the eastern end of the northern parallel ribbon
(Figure~\ref{fig10}(a)). The brightening in the circular ribbon then moves sequentially in a
counterclockwise direction, first to the east and then to the south and then the west. The
brightness reaches the far-southern extreme, opposite the parallel ribbons, at around
$\sim$21:55 UT, with a quasi-circular ribbon visible at around $\sim$22:00 UT
(Figure~\ref{fig10}(c)). The brightness then continues moving further west, forming a
near-circle at around $\sim$22:10 UT (Figure~\ref{fig10}(d)). This circular ribbon was visible
in all of the \textit{SDO}/AIA channels; Figures~\ref{fig11}(a)--(e) respectively show the
circular ribbon in \textit{SDO}/AIA 171, 193, 131, 211, and 304 \AA\ wavelengths. 
Figure~\ref{fig11}(f) shows the circular ribbon overlaid onto an \textit{SDO}/HMI
line-of-sight photospheric magnetogram, with the ribbon in \textit{SDO}/AIA 304 and 1600
\AA\ displayed in green and red contours, respectively. This shows that the circular ribbon is
situated in a negative-polarity region.  

The RHESSI X--ray sources have also been observed during this phase in the region between the parallel ribbons (see Figures~\ref{fig8}(a)--(c)). This suggest that the standard reconnection underneath the erupting flux rope remains ongoing.

The null-point reconnection also heats a ring of new inner fan loops that are visible in  hotter AIA
channels. Figure~\ref{fig12} shows examples of these structures in \textit{SDO}/AIA 335 and 94 \AA\ 
images.  Evidently, the erupting flux rope drives  its enveloping arcade to  reconnect at the
null point in the higher corona, heating this circular ribbon. Figure~\ref{fig14} shows a schematic
for this interpretation, and will be discussed further in Section~\ref{sec5}.  

%------------------------------------------------------------------------------------------

\subsection{Ribbon Timings}
\label{sec4.4}

In reconnection-initiated eruption-onset models, there are (at least) two different mechanisms by
which reconnection can initiate the eruption of a flux rope:  In the first case, the release of
tension force below the flux rope initiates and grows the eruption; this is the runaway
tether-cutting idea \citep{Moore01}.  As we have seen above (Section~\ref{sec2}), it appears that
such  tether cutting produces the parallel flare ribbons under the erupting flux rope in our 
event.  In the second mechanism, it is the removal of  flux overlying the flux rope via
reconnection at a fan null that initiates the flux  rope's eruption; this is the idea of the
magnetic breakout mechanism \citep{Antiochos99}.  In  our event, breakout reconnection evidently
produces the circular ribbon at locations remote from the location from which the flux rope erupts.

As mentioned in Section~\ref{sec1}, a third mechanism that may initiate eruptions is an ideal MHD
instability not involving reconnection.  We do not address this option directly here, in part
because it is difficult to detect the earliest eruptive magnetic motion, since a distinct erupting
filament that can often be used as a tracer of such early magnetic motion is not available in this
event.  We do however observe signatures of both tether cutting and breakout, and so here we
address directly which of these two mechanisms shows the best evidence for initiating the eruption.

To decide which of these two scenarios best fits our event, we use the relative timing of the onset
of the two different flare ribbons.  If the parallel ribbons in the immediate vicinity of the flux
rope brightened prior to the onset of brightening of the remote circular ribbon, that would be an
indication that tether cutting precedes breakout.  Should the order be reversed, that is if the
circular ribbon started prior to the parallel ribbons, this would indicate that breakout precedes
tether cutting.  At least several previous attempts to address this question of tether cutting or
breakout based on relative timing of two-ribbon flare brightening and remote brightening have been
inconclusive, due to insufficient cadence of the observations for the features being studied
\citep[e.g.,][]{moore_et06}.

\textit{SDO}/AIA 1600~\AA\ images show that the parallel ribbons began no latter than 21:38~UT,
and they were essentially fully developed by $\sim$21:50~UT\@.  The circular ribbon on the other
hand did not start until $\sim$21:52~UT, and was not fully developed prior to $\sim$21:55~UT\@.
That is, the null-point reconnection began about 12~min later than the flux rope started to erupt
in step with runaway tether-cutting reconnection and flare heating below (see
Figure~\ref{fig7}(b)).  Therefore, in this case, the evidence unambiguously supports that the
runaway tether-cutting reconnection under the rising flux rope  preceded the breakout reconnection
at the null point.  Hence, in our case we conclude that the tether-cutting reconnection is more
probable than breakout for the initiation of flux-rope eruption (see Figure~\ref{fig5} and
accompanying animation). 

%------------------------------------------------------------------------------------------

\subsection{Accompanying CME}
\label{sec4.5}

This eruption is ejective in that it is accompanied by a CME\@. Figure~\ref{fig13} shows a
SOHO/LASCO  white light coronagraph image of the CME at 01:04:42 UT on 2014 December 19. 
Unfortunately, due to a data gap, only a few images are available and so the CME was not observed
at full cadence. This image shows that the CME accompanying the eruption was a halo CME, with the
outer approximately-circular leading edge of the CME visible in this image. 

%------------------------------------------------------------------------------------------

\section{Schematic and Comparison with Coronal Jet Topology}
\label{sec5}

\begin{figure}
\vspace*{-8cm}
\centerline{
	\hspace*{0.0\textwidth}
	\includegraphics[width=1.3\textwidth,clip=]{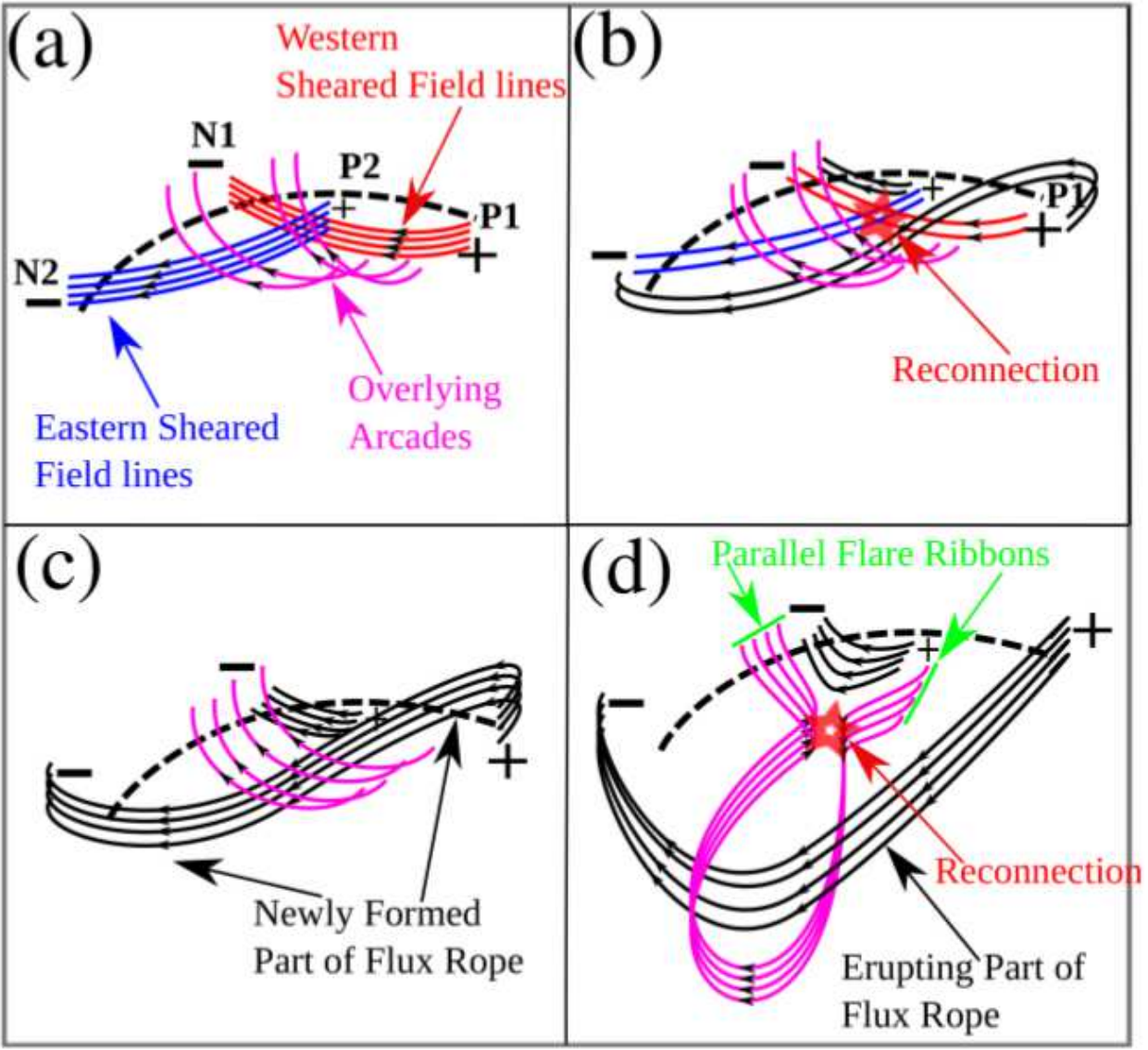}
	}
\vspace*{-6cm}
\caption{Schematic of the flare during the first component of the eruption, which occurs 
in the northeastern lobe of the huge fan dome; this region corresponds to the 
green lines 
in the schematic in Figure~\ref{fig14}. The black dashed line shows the polarity inversion 
line (PIL), the red and blue lines show the impacted sheared field lines that reconnect, the black solid lines represent 
the magnetic lines formed by the reconnection, and the purple lines represent overlying arcade loops.
Red stars in panels (b) and (d) show the reconnection region.}
\label{fig9}
\end{figure}

In our interpretation, this flux rope eruption and accompanying three-ribbon flare followed 
the two-component process outlined in Sections~\ref{sec4.2} and~\ref{sec4.3}. The 
schematic of Figure~\ref{fig9}, discussed above, describes the onset of the two-ribbon component.  
The schematic in Figure~\ref{fig14} describes the overall evolution of the event, including
the circular-ribbon-component null-point reconnection.  

Figures~\ref{fig14}(a) and~(b) show the situation as the flux rope formed in the two-ribbon
component of the event  erupts outward from the surface.  This flux-rope eruption occurs inside
the smaller (i.e.\ more compact) lobe of the inner fan dome.  Because we have learned that the
two-ribbon component of the eruption precedes the circular-ribbon component, we infer that the
circular-ribbon component of the event starts when the erupting flux rope induces reconnection of
its arcade envelope at the null point between the closed (inner) and far-reaching (outer) portions
of the fan structure.  The null-point reconnection results in the circular ribbon when particles
from the null-point reconnection travel along the new dome field lines and impinge upon the
chromosphere, heating that plasma to create the circular flare ribbon (Figure~\ref{fig14}(c)). 
The circular ribbon structure can be seen clearly in several wavelength channels
(Figures~\ref{fig10} and~\ref{fig11}).

Several additional factors support our interpretation outlined in Figure~\ref{fig14}: (1) the 
existence of null-point-like structure in the PFSS extrapolation; (2) the observation that the
eruption of the flux rope is directed towards the null point; (3) the formation of the circular
ribbon; and (4) the bserved RHESSI X--ray sources below the erupting flux rope, which is consistent with the geometry of
Figure~\ref{fig14}.  Thus the erupting flux rope plays a key role in driving both the standard
flare  reconnection \citep[e.g.,][]{Liu07} that results in the parallel ribbons, and the
null-point breakout reconnection \citep[e.g.,][]{Sun13,Joshi15} that results in the circular ribbon.

The schematic in Figure~\ref{fig14} for our event is essentially the same as the schematic in
\citet{Sterling15} for X-ray jets in polar coronal holes \citep[and similar schematics for  active
region coronal jets in][]{sterling_et16,sterling_et17}. The field configuration we observe for the
large-scale three-ribbon event studied here has a base size of $\sim$10$^5$~km, as determined by
the approximate diameter of the circular ribbon.  For the coronal jets, the same configuration
occurs on a substantially smaller size scale, with a base size of $\sim$10$^4$~km.  

Other than size scale, the eruption in our event has two main differences from the jet eruptions
of \citet{Sterling15}.  First, the erupting flux rope in our event, as pointed out in
Section~\ref{sec4.2}, carried no trackable dark filament of cool plasma, whereas the erupting flux
rope in each of the jets studied by \citet{Sterling15} did carry a trackable minifilament of cool
plasma.  Second, the erupting sheared-core arcade with the erupting flux rope in its core in our
event blows out into the high corona and solar wind as a CME (that is, in this case most of the
erupting arcade is not opened by the breakout reconnection), whereas in the coronal jets of
\citet{Sterling15} that are blowout jets, all of the blown-out arcade, including its
minifilament-carrying core, is opened by the null-point reconnection and becomes part of the spire
of the jet.

Figure~\ref{fig14} shows a variation of the jet schematic given in \citet{sterling_et16} where the
cool minifilament, which would reside inside of the curled field lines on the  left-hand-side
bipole in panel~(a), is omitted; thus, in this case the curled field lines indicate a
minifilament-flux rope field that does not carry cool material.  This is a better match than the
schematic in \citet{sterling_et16} for the situation studied here with the large-scale eruption
that does not include an obvious erupting  cool filament.  Even in the case of coronal jets though,
even without the minifilament the minifilament-flux rope field could erupt, just as  the
large-scale event studied here had an erupting flux rope that did not carry a filament.  Such
minifilament flux ropes without cool minifilament material may erupt to cause jets in some cases
\citep[][]{sterling_et17}.

In Figure~\ref{fig14}(b), reconnection is occurring beneath the erupting minifilament field,
producing a brightening at the jet's base (jet-base bright point, or JBP); this corresponds to the
two-ribbon flare that we observe in the large-scale event here.  Figure~\ref{fig14}(c) shows the
arcade envelope of the erupting minifilament during reconnection at the null, and this makes new
hot loops on top of  the larger bipole.  The jet spire initially forms at the location of the null,
and it can spread horizontally away from the JBP depending on the subsequent evolution of the
erupting flux rope.  

Comparing this jet picture with the large-scale eruption studied here, the JBP located between
locations A and B of Figure~\ref{fig14}(c) corresponds to the two-ribbon flare of
Figure~\ref{fig9}, and the large loop between locations B and C of Figure~\ref{fig14}(c)
corresponds to the illuminated inner fan dome.  Comparing the jet
schematics with the large-scale eruption in Figures~\ref{fig12},~\ref{fig14}(c), and~\ref{fig14}(d) shows well how the two circumstances
are analogous: The JBP of jet is corresponds to the bright flare loops on the north
side of the erupting region in Figures~\ref{fig12}, and~\ref{fig14}(c), and the new (red) field line over the larger
lobe in  Figure~\ref{fig14}(c) corresponds to the heated and illuminated inner-fan loops of
Figure~\ref{fig12}.  

Because \citet{Sterling15} observed jets near the limb, remote flare brightenings in the form of a
circular ribbon were not well situated for visual detection.  Some studies of on-disk jets in
active regions however observed such circular ribbons
\citep[e.g.,][]{Wang12,wang_et14,jiang_et15,sterling_et16,zhang_et16,li_et17}.  Similar circular
ribbons are also apparent, but often with lower intensity, in quiet Sun jets \citep{panesar_et16}. 
Partial circles of illumination at the bases of surges have also been observed \citep[e.g.,][]{ohman72,Wang12,sterling_et16,Hong17}.

According to the schematic of \citet{Sterling15}  (Figure~2 of that paper), a variation of which is
the Figure~\ref{fig14} schematic, the implication is that in jets the JBP-producing flare
reconnection occurs prior to the null-point reconnection, which in Figure~\ref{fig14} does not
occur until panel~(c).  Actually however, whether the JBP-producing reconnection or the null-point
reconnection occurs first in jets is so far an unexplored question.  Our work here in
Section~\ref{sec4.4} indicates that, at least for the large-scale eruptions presented here, the
flare reconnection precedes the null-point reconnection. 

Regarding the longer-term evolution, in the case of the jet, the erupting minifilament-flux-rope
field moves out in a columnated fashion to form the jet.  Typically however the erupting
minifilament flux rope completely reconnects with the surrounding open (or far-reaching) field, and
thus tends to have a highly-columnated jet structure.   In the large-scale eruption, the flux rope
contains enough flux so that the reconnection with the surrounding field does not consume much of
the flux rope, and so the erupting arcade with the flux rope in it largely escapes intact to become
the CME with a flux-rope core.  

Therefore, the magnetic conditions resulting in the double-ribbon eruption of 2014 December~18
studied here happen across the wide range of size scales of solar eruptive events.

%------------------------------------------------------------------------------------------

\section{Discussion}
\label{sec6}

We have presented a detailed analysis of an ejective eruption accompanying a \textit{GOES} M6.9
class flare, using multi-wavelength data sets from \textit{SDO}/AIA, \textit{RHESSI} and
\textit{GOES}\@. The eruption produced three flare ribbons: two parallel ribbons that form
according to  the standard eruption/flare model as a flux rope forms and erupts, and a circular
ribbon resulting from null-point reconnection between the arcade envelope of that  erupting flux
rope with a fan canopy of open field. The main results of the work are as follows:

\begin{enumerate}

\item The observations as well as the PFSS extrapolation calculations clearly show the existence
of a large-scale fan-spine type magnetic configuration over and around the eruption region.  The
northeast side of the inner fan lobe contains a strong neutral line on its inside.

\item The whole event has two aspects, or components: The ``two-ribbon'' component starts along
the strong neutral line of the closed inner fan dome, and produces a flux rope and flare with
parallel flare ribbons in accordance with the standard model for eruptions. The ``circular-ribbon"
component occurs when the enveloping arcade of the erupting  flux rope undergoes reconnection at
the null point of the overlying fan; this results in formation of a quasi-circular flare ribbon at
the locations where the newly-closed fan field is rooted in the lower atmosphere.  

\item The erupting flux rope therefore drives the production of all three flare ribbons.  Standard
flare tether-cutting reconnection is responsible for the formation of the two parallel ribbons,
while the  null-point reconnection is responsible for the large-scale quasi-circular ribbon.  

\item Leading to the eruption onset, flux cancelation occurs along the main PIL from which the
flux rope erupts, and at the location where the initial flare brightenings (pre-flare
brightenings) occur. This is consistent with tether cutting from evolutionary
photospherically-driven flux cancelation at the PIL triggering the eruption, and with runaway
tether-cutting reconnection initiating the eruption and growing the flux rope.

\item A timing analysis shows that in this event the parallel ribbons unambiguously brightened 
prior to the onset of brightening of the circular ribbon.  This is consistent with runaway
tether-cutting reconnection starting prior to the start of the null-point ``breakout'' reconnection
at the fan-dome magnetic null.

\item The observed magnetic topology and overall morphology is consistent with this eruption being 
a large-scale version of the same processes occurring in coronal jets that occur according to the minifilament-eruption picture.

\end{enumerate}

There have been previous observational studies of eruptions resulting in circular ribbons, based
on magnetic configurations similar to that we study here: a fan dome with an erupting bipolar
region  inside it.  These studies encompass both relatively-small-scale \citep{Sun13,Jiang13}, as
well as relatively-large-scale \citep{Joshi15} eruptions. \cite{Joshi15} investigated a
configuration similar to that here in a different large-scale eruption. The observed results in this eruptive event confirms the earlier proposed scenario by \cite{Joshi15} for a similar kind of eruption and multiple ribbons flare. These studies show that
the eruptions inside of fan structures is a common occurrence on the Sun.  Remote flare
brightenings (analogous to the circular ribbons observed here) and transient fan-like-loop heating
can also occur in somewhat different magnetic geometries, such as when the arcade envelope of an
erupting flux rope rams into nearby coronal hole field \citep{sterling_et01}.

In addition to studies suggesting that a flux rope (or sigmoid) eruption from the base of the fan
dome drives null-point reconnection, and hence produces the circular ribbon flare
\citep{Sun13,Jiang14,Joshi15}, other studies suggest that shearing motion of the feet of the closed
fan field by photospheric flow may drive the reconnection at the null point \citep{Vemareddy14b}. In
our case here though it is very clear that the eruption of flux rope drives the null-point
reconnection (see Section~\ref{sec4.2}). 

\begin{figure}
\vspace*{-8cm}
\centerline{
	\hspace*{0.0\textwidth}
	\includegraphics[width=1.4\textwidth,clip=]{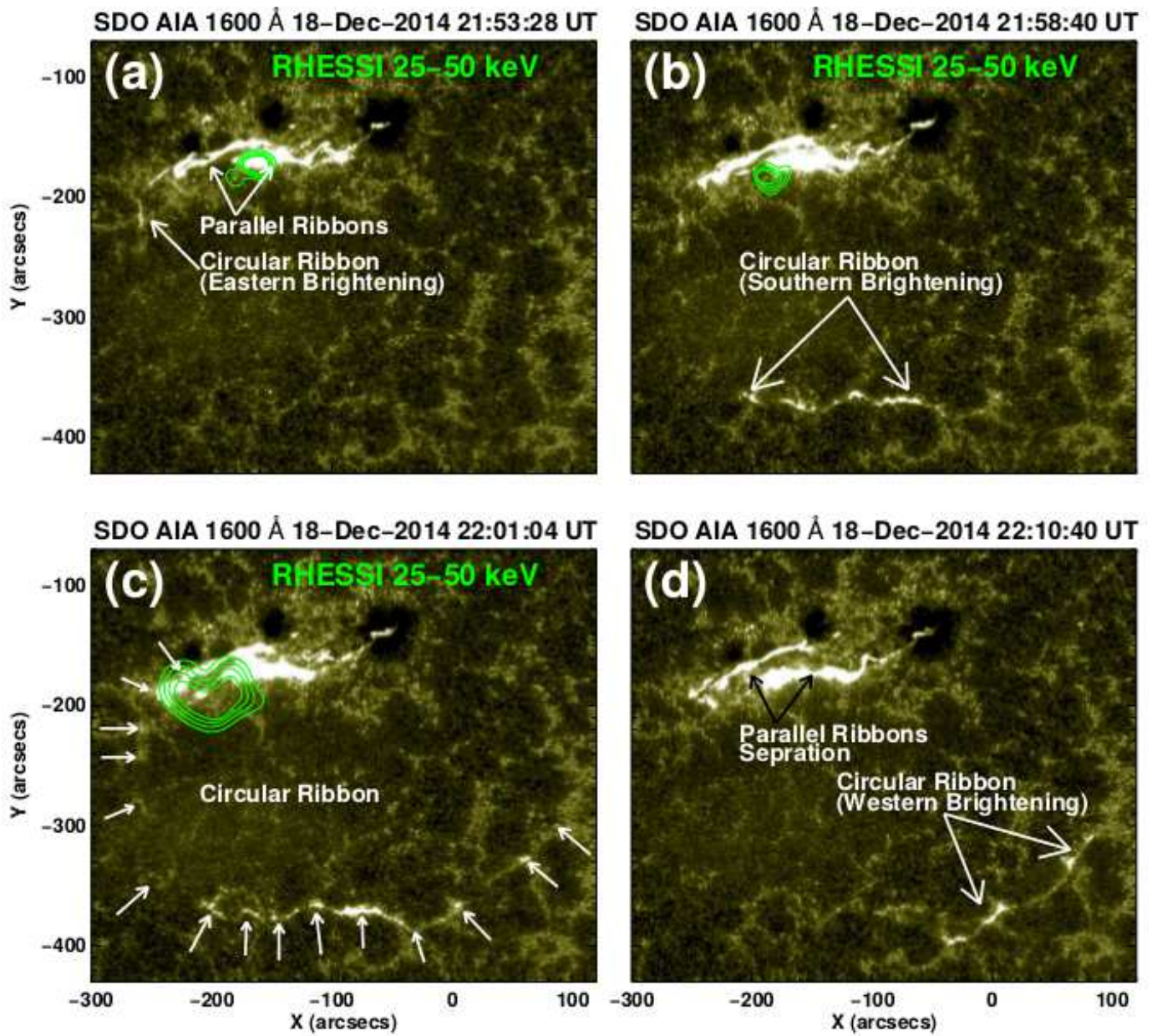}
	}
\vspace*{-7cm}
\caption{Sequence of selected \textit{SDO}/AIA 1600 \AA\ images, showing the evolution of the 
two-ribbon parallel flare ribbons and the circular ribbon. The green contours in panels (a) and (b) map
\textit{RHESSI}  X-ray intensity in the energy range 25-50 keV, generated with the PIXON algorithm. The
contours  levels are 20\%, 30\%, 40\%, and 50\% of the peak intensity.
The  integration time is 20~s.}
\label{fig10}
\end{figure}

\begin{figure}
\vspace*{-8cm}
\centerline{
	\hspace*{0.0\textwidth}
	\includegraphics[width=1.2\textwidth,clip=]{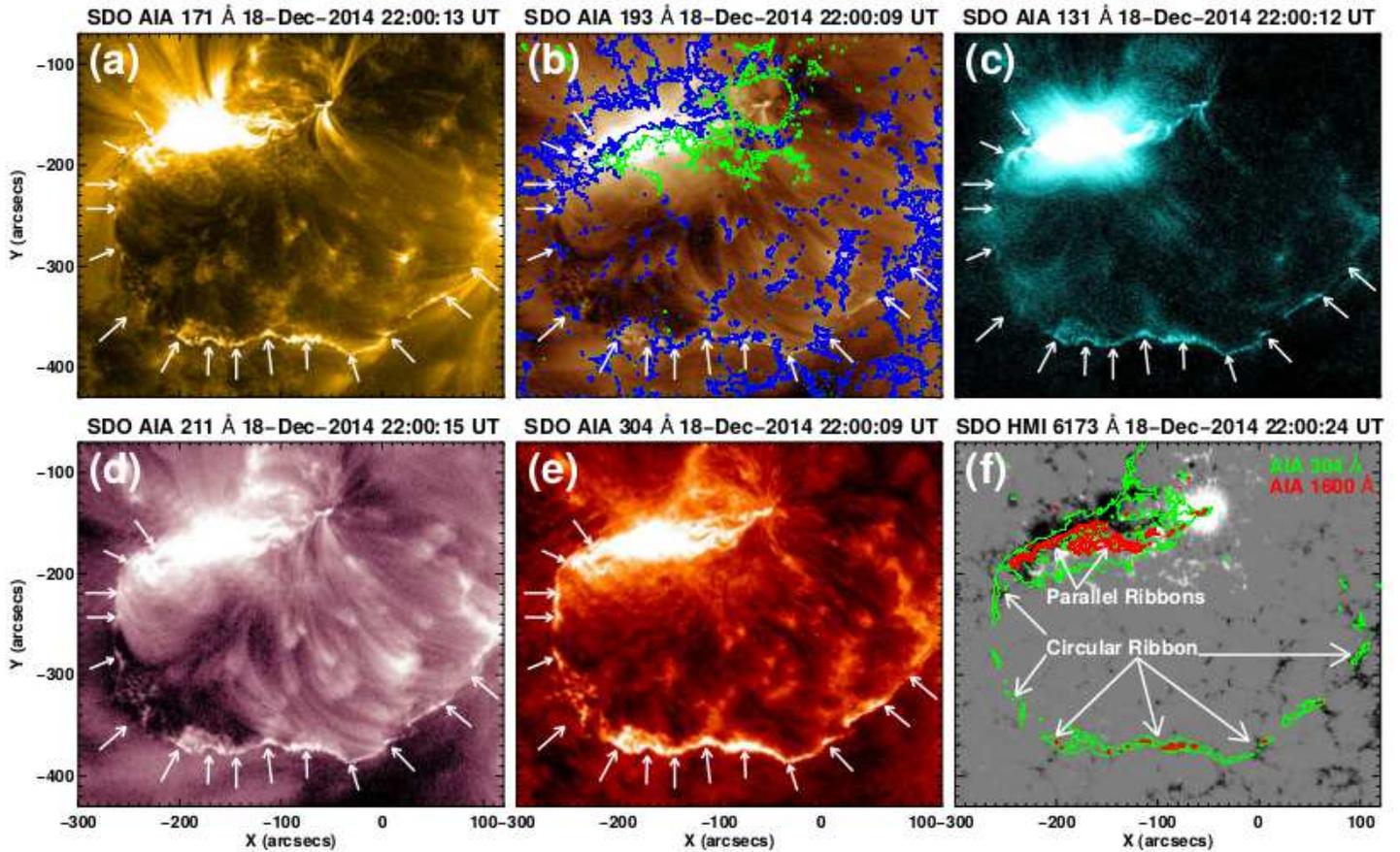}
	}
\vspace*{-7cm}
\caption{\textit{SDO}/AIA 171 (a), 193 (b), 131 (c), 211 (d), and 304 (e) \AA\ images at $\sim$22:00 UT, showing the circular ribbon. In panel (b) the blue and green contours show the photospheric magnetic field contours for negative and
positive polarity field, respectively, with contour levels of $\pm40,\pm50, \rm{and } \pm100$ Gauss. (f) \textit{SDO}/HMI line-of-sight magnetogram at $\sim$22:00 UT, with  overlaid \textit{SDO}/AIA 1600~\AA\ (red) and 304~\AA\ (green) contours. The contour levels are 5\%, 10\%, 20\%, and 30\% of the peak intensities.}
\label{fig11}
\end{figure}

\begin{figure}
\vspace*{-8cm}
\centerline{
	\hspace*{0\textwidth}
	\includegraphics[width=1.2\textwidth,clip=]{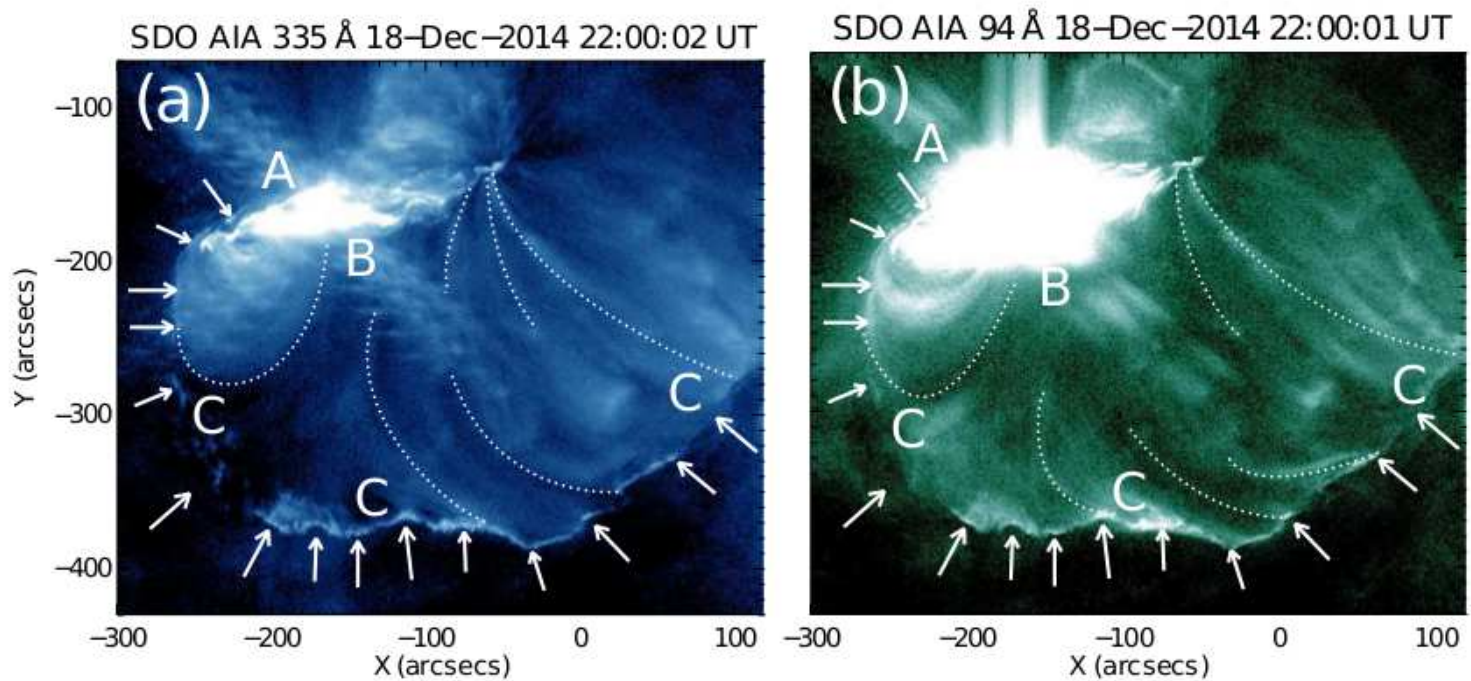}
	}
\vspace*{-8cm}
\caption{\textit{SDO}/AIA 335 and 94 \AA\ images at $\sim$22:00 UT showing 
inner-fan-dome new closed loops made and transiently heated by the null-point reconnection 
driven by the erupting arcade, which was blown out by the erupting flux rope in its core. 
The dotted lines show example field lines, and the arrows show
the location of the circular flare ribbon, which forms at the footpoints of the 
transiently-heated portion of the inner-dome field.} 
\label{fig12}
\end{figure}

\begin{figure}
\vspace*{-10cm}
\centerline{
	\hspace*{0.0\textwidth}
	\includegraphics[width=1.4\textwidth,clip=]{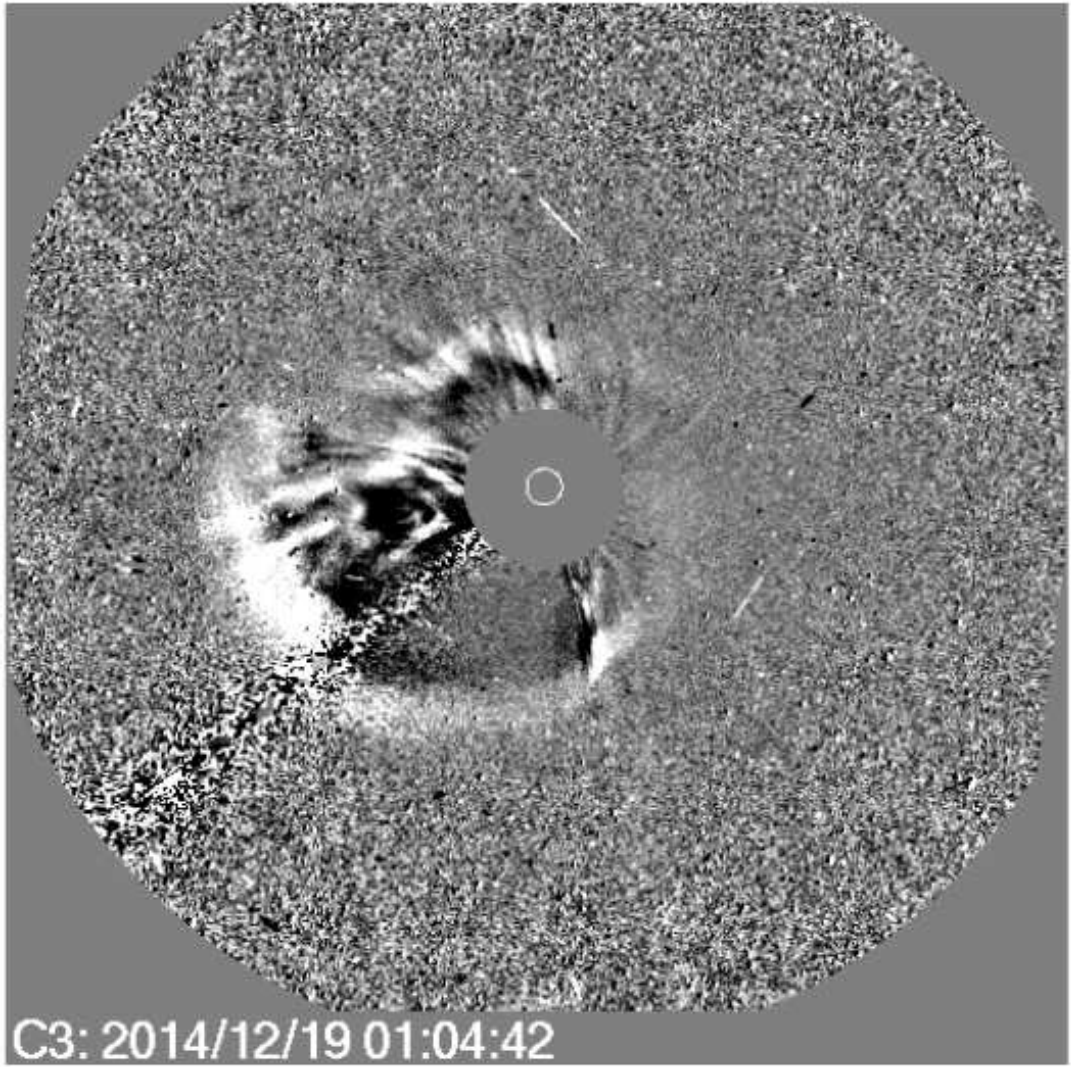}
	}
\vspace*{-7cm}
\caption{SOHO/LASCO C3 coronagraph image at 01:04:42 UT, showing the halo CME produced by the eruptive event studied here. The difference image is collected from \textit{SOHO}/LASCO catalog (\url{http://cdaw.gsfc.nasa.gov/CME\_list/}}
\label{fig13}
\end{figure}

\begin{figure}
\vspace*{-10cm}
\centerline{
	\hspace*{0.0\textwidth}
	\includegraphics[width=1.4\textwidth,clip=]{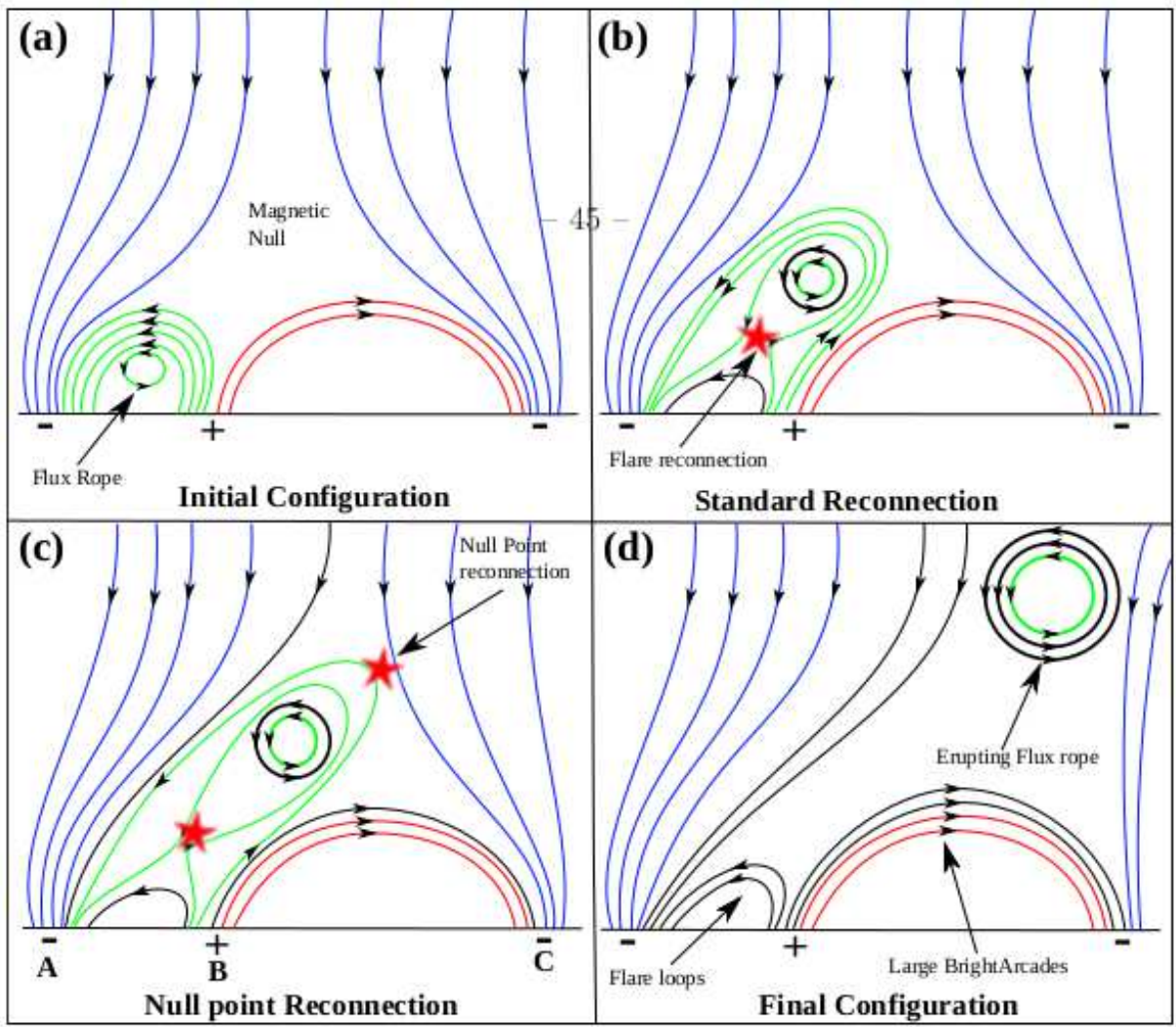}
	}
\vspace*{-7.5cm}
\caption{\footnotesize 
Schematic for our large-scale event studied here. Blue lines and open-ended black lines 
represent field that reaches distant solar locations, or is open. Red and green lines represent 
the initially closed field lines of the inner shell of the fan dome. Black lines represent field lines 
formed after reconnection. Red stars mark the reconnection locations. This eruption schematic
matches very well with both the large-scale-eruption scenario proposed by \cite{Joshi15} for
a different large-scale eruption, and the jet-formation schematic presented in jet papers by 
\cite{Sterling15,sterling_et16,sterling_et17}.  Panel (a) represents the initial configuration.  From the perspective 
of the 2014 December 18 eruption of this paper, the green field on the left-hand side erupts to 
produce the two-ribbon flare in the two-ribbon component of the event (Figure~\ref{fig9}), with the ribbons 
forming, between locations A and B (see labels in panel C), due to tether-cutting reconnection.  
The right-hand-side reconnection corresponds to the null-point-reconnection component of the 
event, with the newly-formed large-lobe loops between locations B and C corresponding to the 
new loops made and heated by the null-point reconnection (see Figure~\ref{fig12}), and the foot of that 
bright loop at location C corresponding to one location of the circular ribbon as viewed on a 
2D cross sectional plane.  From the jet perspective:  Panels (b) and (c) show a miniature flux 
rope erupting, with the tether-cutting reconnection (called “internal reconnection” in the Sterling 
et al. jet papers) on the left-hand side (panels b and c) occurring to make a bright structure 
between locations A and B in (c). This is what Sterling et al. call a “jet-base bright point” (JBP;  
a miniature flare arcade having a miniature pair of flare ribbons in its feet in the case of jets).  
Thus the JBP is a miniature flare, corresponding to the two-ribbon flare of the large-scale event. 
The null-point reconnection (also called “breakout reconnection” or “interchange reconnection,” 
and called “external reconnection” in the Sterling et al. jet papers) on the right-hand side (panel 
(c)) creates new bright closed loops between locations B and C, and also new far-reaching field 
lines. This reconnection of the far-reaching field results in a heated jet column in the Sterling 
et al. jet papers.}
\label{fig14}
\end{figure}

Also however, the sequential circular-ribbon brightening from east to west (see Figure~\ref{fig10}
and  accompanying animations) may provide some evidence of slip-running type reconnection at the
null point.  Similar sequential ribbon brightening has previously been observed and discussed
\citep{Masson09,Reid12,Sun13}. 

The eruption of flux ropes to produce both a set of standard-flare parallel ribbons and a circular
ribbon  has only recently started to be studied in detail; our study here provides another example
of such an event.  These studies, together with those of jets and other phenomena, are helping to
put together a coherent picture of various solar eruptions.  Observational studies such as these
promise to provide important input for numerical simulations of eruptions in various magnetic
geometries. More observational studies  as well as simulations are needed to refine the
understanding of such events. Moreover, the study of events such as the one here occurring near
solar-disk center will provide important inputs for improving space weather forecasting.

%------------------------------------------------------------------------------------------

\acknowledgments

The authors thank referee for their comments and suggestions. We thank the \textit{SDO}/AIA, \textit{SDO}/HMI, \textit{GOES}, \textit{SOHO}/LASCO and \textit{RHESSI} teams for
providing their data for the present study. This work is supported by the BK21 plus program
through the National Research Foundation (NRF) funded by the Ministry of Education of Korea. NCJ
thanks the School of Space Research, Kyung Hee University for providing a Postdoctoral grant.
A.C.S. and R.L.M. were supported by funding from the Heliophysics Division of NASA's Science
Mission Directorate through the Heliophysics Guest Investigators (HGI) Program.

%------------------------------------------------------------------------------------------
%\bibliography{reference.bib}
%\bibliographystyle{apj} % style aa.bst
%\bibliography{reference} % your references Yourfile.bib
%\bibliography{reference} % your references Yourfile.bib

%------------------------------------------------------------------------------------------

%\clearpage

%------------------------------------------------------------------------------------------

\clearpage

%------------------------------------------------------------------------------------------

\clearpage

%------------------------------------------------------------------------------------------

\clearpage

%------------------------------------------------------------------------------------------

\clearpage

%------------------------------------------------------------------------------------------

\clearpage

%------------------------------------------------------------------------------------

\clearpage

%------------------------------------------------------------------------------------------

\clearpage

%------------------------------------------------------------------------------------------

\clearpage

%------------------------------------------------------------------------------------------

\clearpage

%------------------------------------------------------------------------------------------

\clearpage

%------------------------------------------------------------------------------------------

\clearpage

%------------------------------------------------------------------------------------------

\clearpage

%------------------------------------------------------------------------------------------

\clearpage

%------------------------------------------------------------------------------------------

\end {document}